\documentclass[aps, prb, twocolumn, showpacs, 10pt, floatfix, superscriptaddress, floatfix, longbibliography]{revtex4-2}

\usepackage{setspace}

\usepackage[english]{babel}
\usepackage[utf8]{inputenc}
\usepackage[T1]{fontenc}

\usepackage{amsmath}
\usepackage{braket}
\usepackage{amssymb}
\usepackage[normalem]{ulem}
\usepackage{dsfont}
\usepackage{bm}
\usepackage[dvipsnames]{xcolor}
\usepackage{stmaryrd}
\usepackage{bbm}
\usepackage{mathtools}

\usepackage{multirow}
\usepackage{enumitem}

\usepackage{color}
\usepackage[dvipsnames]{xcolor}

\usepackage[most]{tcolorbox}
\newtcbox{\othermathbox}[1][]{nobeforeafter, math upper, tcbox raise base, enhanced, rounded corners, colback=black!5, colframe=black, left=0.3em, top=0.3em, right=0.3em, bottom=0.4em}
\usepackage{empheq}
\usepackage[colorlinks,allcolors=blue]{hyperref}

\makeatletter
\def\l@subsection#1#2{}
\def\l@subsubsection#1#2{}
\makeatother

\newcommand{\beq}{\begin{equation}}
\newcommand{\eeq}{\end{equation}}
\newcommand{\bA}{\textbf A}
\newcommand{\bB}{\textbf B}

\newcommand{\bp}{\textbf p}

\newcommand{\bx}{\textbf x}
\newcommand{\by}{\textbf y}
\newcommand{\br}{\textbf r}
\newcommand{\bs}{\textbf s}

\newcommand{\cO}{\mathcal{O}}

\newcommand{\ie}{{\it i.e.}\ }
\newcommand{\eg}{{\it e.g.}\ }
\newcommand{\cf}{{\it cf.}\ }

\newcommand{\RR}{\mathbb{R}}

\def\Var{\operatorname{Var}}

\def\ii{\mathrm{i}}
\def\ee{\mathrm{e}}
\def\dd{\mathrm{d}}

\def\action{\mathcal{A}}
\def\Vnull{V_0}
\def\droplet{\Omega}
\def\symplectic{F}
\def\conformal{G}
\def\greensFunc{K}
\def\LSFunc{U}
\def\LSOp{\hat{U}}

\definecolor{MyRed}{RGB}{220,60,10}
\definecolor{LightYellow}{RGB}{255,245,108}
\definecolor{LightBrown}{RGB}{255,188,0}
\definecolor{MiddleBrown}{RGB}{199,146,0}
\definecolor{DarkBrown}{RGB}{143,104,0}
\definecolor{DarkerBrown}{RGB}{87,62,0}
\definecolor{Purple}{RGB}{255,0,188}

\begin{document}

\title{Quantum Hall edges beyond the plasma analogy}

\author{Per Moosavi}
\email{per.moosavi@fysik.su.se}
\affiliation{Department of Physics, Stockholm University, 10691 Stockholm, Sweden}

\author{Blagoje Oblak}
\email{oblak@math.univ-lyon1.fr}
\affiliation{Universit\'e Claude Bernard Lyon 1, ICJ UMR5208, CNRS, 69622 Villeurbanne, France}

\author{Bastien Lapierre}
\email{blapierre@princeton.edu}
\affiliation{Department of Physics, Princeton University, Princeton, New Jersey 08544, USA}

\author{Benoit Estienne}
\email{estienne@lpthe.jussieu.fr}
\affiliation{Sorbonne Universit\'e, CNRS, Laboratoire de Physique Th\'eorique et Hautes Energies, LPTHE, F-75005 Paris, France}

\author{Jean-Marie St\'ephan}
\email{jean-marie.stephan@ens-lyon.fr}
\affiliation{Universit\'e Claude Bernard Lyon 1, ICJ UMR5208, CNRS, 69622 Villeurbanne, France}
\affiliation{ENS de Lyon, CNRS, Laboratoire de Physique, F-69342 Lyon, France}

\date{May 14, 2025}

\begin{abstract}
We demonstrate that the widely used plasma analogy is unreliable at predicting edge properties of quantum Hall states. This discrepancy arises from a fundamental difference between quantum Hall droplets and plasmas (Coulomb gases): the former are incompressible liquids subject to area-preserving deformations, while the latter are governed by electrostatics and thus involve conformal transformations. Consequently, the plasma analogy fails at the edge, except in fine-tuned geometries, as it does not account for the emergent local edge velocity. We quantitatively show how the analogy's failure affects physical quantities, such as fluctuations of local observables and absorption rates in microwave spectroscopy, measurable in both solid-state experiments and quantum simulators.
\end{abstract}

\maketitle

\section{Introduction}

This paper critically contrasts two commonly related physical systems. The first is a topological phase of matter famously responsible for the quantum Hall (QH) effect \cite{Klitzing, Tsui, TKNN, LaughlinQuantized, Laughlin:1983anomalous}: a two-dimensional (2D) incompressible quantum liquid of electrons in a strong perpendicular magnetic field. The second is a 2D Coulomb gas (CG) \cite{Jancovici:1981, Jancovici:1993, Jancovici:1995, ForresterJancovici:1996, DiFrancesco:1994, WiegmannZabrodin:2003a, LebleSerfaty:2018, Jancovici:1982ii, AlastueyJancovici:1984, ChoquardEtAl:1987}: a classical statistical system of planar point charges interacting via a 2D electrostatic potential. Both systems are confined into droplets of finite area by trapping potentials, inherently present in any experiment.

Our investigation is motivated by what is known as the \emph{plasma analogy}, stating that QH systems are equivalent to suitable CGs. This correspondence underlies Laughlin's theory of the fractional QH effect \cite{Laughlin:1983anomalous} and provides a handle on microscopic aspects that are otherwise out of reach due to strong electronic correlations. The analogy naturally also holds for integer QH droplets of free electrons in harmonic potentials, whose one-body wave functions coincide with those of a random matrix ensemble---the (elliptic) Ginibre ensemble \cite{Ginibre:1965, Girko:1986}. This includes the simplest, familiar case of isotropic QH droplets, whose wave functions have definite angular momentum. However, the plasma analogy remains untested for arbitrary potentials. This is a significant gap in the literature, given that realistic QH potentials are always disordered and anisotropic \cite{Halperin, Champel1, Champel2}. Additionally, even in quantum simulators, where the potential can be controlled \cite{FletcherEtAl:2021, Mukherjee:2021jjl, YaoEtAl:2024, Crepel:2023kbu, Leonard:2023ndq, Binanti:2023ozm, Braun:2024yjh}, it is desirable to know how much of the plasma analogy may be harnessed to model QH physics.

In the present paper, we put the plasma analogy to the test. This relies on a recent semiclassical treatment of integer QH droplets in general anisotropic potentials \cite{OLMSE:2024}, allowing for direct comparisons with CG predictions. The question is: Does the plasma analogy work for generic QH droplets? Strikingly, while reproducing bulk properties, the answer for edge phenomena is negative: QH droplets and CGs of identical shape almost always have different electronic correlations along the edge, resulting \eg in unequal values for fluctuations of local observables. Such fluctuations can be measured, at least indirectly, through microwave absorption experiments \cite{Talyanskii, Andreev, Cano, Mahoney, FrigerioEtAl:2024}, so the implications of the QH/CG mismatch are both theoretical and experimental.

\begin{figure}[t]
\includegraphics[width=.4\textwidth]{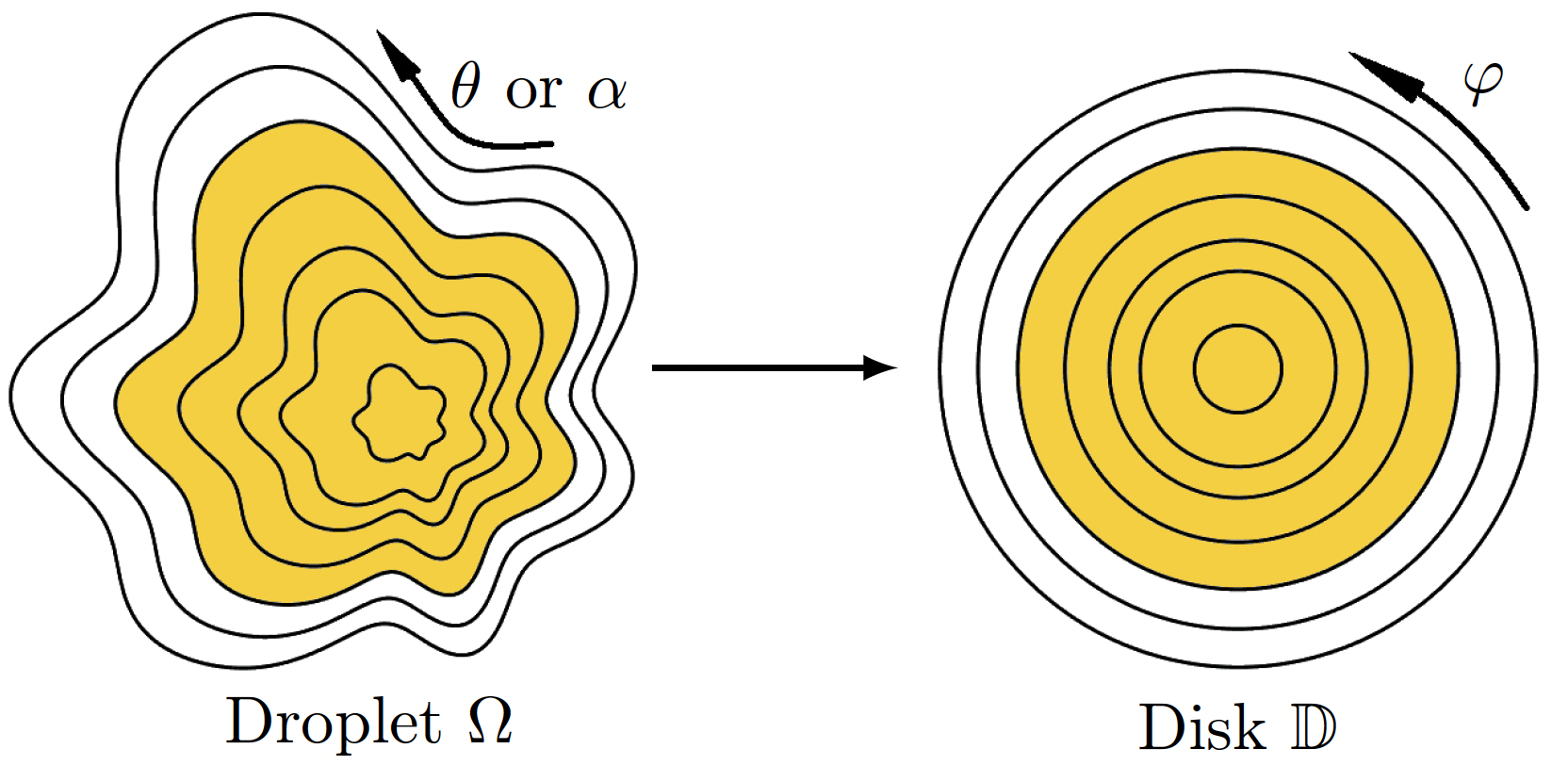}
\caption{QH droplets and CGs, shown in yellow along with a few level curves of the confining potential in the QH case. Both involve maps that send them on a disk $\mathbb{D}$, but with different geometric constraints, namely symplectic (area-preserving) in QH and conformal (angle-preserving) in CG. Given the same initial droplet $\droplet$, such maps do not generally coincide, leading to different angle coordinates governing the emergent long-range edge correlations: the coordinate is $\theta$ for QH, $\alpha$ for CG, and $\varphi$ in the isotropic case.}
\label{fig1}
\end{figure}

These differences between QH droplets and CGs should not come as a surprise. Since CGs ultimately rely on electrostatics, their physics is entirely dictated by their shape---albeit in a nonlocal manner. By contrast, QH physics is that of an incompressible fluid, so it depends not only on a droplet's shape but also on the local slope of the potential at the edge. This is the core of our argument. Abstractly, it can be phrased in terms of maps that send anisotropic droplets on their isotropic cousins, as in Fig.~\ref{fig1}: the map is \emph{symplectic} (area-preserving) for QH, but \emph{conformal} (angle-preserving) for CG. It is still true that QH and CG predictions sometimes coincide, but this only occurs for fine-tuned setups, such as harmonic potentials \cite{ForresterJancovici:1996}. Needless to say, no such fine-tuning is expected in solid-state experiments, where even just determining the confining potential is a challenge.

The paper is organized as follows. Sec.~\ref{se2} briefly reviews the links between QH physics and symplectic geometry on one hand, and CGs and conformal geometry on the other. This highlights the QH/CG discrepancy, whose implications are then examined in the remaining sections. Indeed, Sec.~\ref{se3} illustrates the difference between QH and CG edge correlations through various explicit examples, and Sec.~\ref{se5} focuses on the consequences for fluctuations of local observables, some of which are evaluated using Monte Carlo numerics on the CG side. Finally, we conclude in Sec.~\ref{se6}. Further details on the mismatch, including brief reviews of QH droplets and CGs, are deferred to the Supplemental Material (SM) \cite{SM}.

\section{Quantum Hall is symplectic, Coulomb gas is conformal}
\label{se2}

Here, we review the basics of QH and CG physics and outline their respective links with symplectic and conformal geometries.

\medskip
\noindent\textbf{Quantum Hall droplets.}
To begin, consider a classical particle with charge $q$ in the plane, subjected to a uniform perpendicular magnetic field $\bB$ and a confining potential $V$. Assuming the magnetic field is strong, the particle traces high-frequency cyclotron orbits whose guiding center $\bx(t)$ slowly drifts along equipotentials (level curves) of $V(\bx)$. The drift velocity is
\beq
\label{ss1q}
\dot x^i
=
-\epsilon^{ij}\frac{\partial_jV}{qB}
\eeq
in Cartesian coordinates $\bx = (x^1, x^2) = (x, y)$, using the Levi-Civita symbol $\epsilon^{ij}$. Viewing Eq.~\eqref{ss1q} as an equation of motion for the coordinates $(x,y)$ with Hamiltonian $V(x,y)$, the space where the particle lives effectively becomes the phase space of its guiding center. Thus, guiding-center coordinates do not commute; they are instead canonically conjugate, satisfying $\{x,y\} = 1/qB$.

This symplectic structure has important implications. Namely, the guiding-center velocity \eqref{ss1q} is tangent to equipotentials of $V$, and its norm
\beq
\label{v_theta}
v(\bx)
\equiv
\sqrt{\dot{x}^2 + \dot{y}^2}
=
\frac{1}{qB} \big|\nabla V(\bx)\big|
\eeq
is generally nonconstant. One can nevertheless make the motion look isotropic in terms of well-chosen coordinates, at least when the equipotential has the topology of a circle. Indeed, let the equipotential have length $L$ and label its points by the arc length $s$ to define an angle
\beq
\label{sa}
\theta(s)
\equiv
2\pi
\int_0^s\frac{\dd s'}{v(s')}
\,\,\bigg/
\int_0^L\frac{\dd s'}{v(s')},
\eeq
where the lower integration limit is arbitrary and $v(s)$ is the norm \eqref{v_theta} at $s$. This is the \emph{canonical angle coordinate} along the equipotential, unique up to rotation. It is conjugate to a \emph{canonical action coordinate} $\action$, with dimensions of area, such that $\dd\action \wedge \dd\theta = 2\pi\,\dd x\wedge\dd y$ \footnote{In Hamiltonian mechanics, $\action$ measures the phase-space area enclosed by an energy level. Here, phase space is the noncommutative space of guiding-center positions, so \emph{action} and \emph{area} are synonymous.}. Each value of $\action$ labels a level curve of $V$, so the potential can be written as $V(\bx) = \Vnull(\action(\bx))$ for a strictly increasing function $\Vnull(\cdot)$ of $\action$ only, at least in a neighborhood of the equipotential \footnote{Our normalization of the action coordinate $\action$ differs from that of the action coordinate $K$ in \cite{OLMSE:2024}. The two are related by $\action = 2\pi\ell^2 K$. Relatedly, $\Vnull(\cdot)$ here differs from that in \cite{OLMSE:2024} by a factor $2\pi$ in the argument.}. These coordinates make the guiding center motion in Eq.~\eqref{ss1q} uniform in the sense that
\beq
\dot \action = 0,
\qquad
\dot{\theta} = \frac{2\pi}{qB} \Vnull'(\action) > 0.
\label{s15s}
\eeq
When $V$ has a unique minimum surrounded by nested level curves, action-angle coordinates are globally well defined on the plane, and $\action$ can be chosen to be the actual area enclosed by an equipotential. For example, if $V$ is isotropic, then $\action = \pi r^2$ and $\theta = \varphi$ in polar coordinates.

The problem of solving the equation of motion \eqref{ss1q} thus boils down to finding a canonical (area-preserving) map
\beq
\symplectic:
\RR^2 \to \RR^2:
(x,y) \mapsto (\action,\theta)
\label{t15s}
\eeq
that converts the anisotropic potential $V$ into its isotropic cousin $\Vnull(\action)$, as in Fig.~\ref{fig1}. Here, we write the map as if it was globally well defined, which is generally neither true nor required: all one needs is that it sends a neighborhood of an equipotential on a neighborhood of a circle in a smooth, bijective, and area-preserving manner. In our case, the equipotential of interest will be the edge of a droplet, whose topology is always that of a circle no matter how disordered the bulk potential is.

Turning to quantum physics, the corresponding problem of an electron in a strong magnetic field with a harmonic potential dates back to \cite{Fock:1928, Darwin:1931} and is solvable using special functions. More generally, for any confining potential $V$, the above classical structures carry over to the quantum case, including many-body states at low energy. Position operators projected to the lowest Landau level (LLL) fail to commute, and the position of a coherent state in the LLL satisfies Eq.~\eqref{ss1q} in the semiclassical limit of weak potentials and strong magnetic fields \cite{Trugman, Joynt, Champel4}. Relatedly, semiclassical eigenstates in the LLL are localized on equipotentials, each enclosing a quantized area $\action = 2\pi m\ell^2$, where $\ell \equiv \sqrt{\hbar/|qB|}$ is the magnetic length and $m$ is a (large) positive integer \cite{Charles:2003a, Charles:2003b, OLMSE:2024}. Extrapolating to a many-body system of $N\gg1$ free electrons in the LLL is straightforward: they occupy the domain $\droplet$ where $V(\bx) \leq E_{\textrm{F}}$ for some Fermi energy $E_{\textrm{F}}$, forming a \emph{quantum Hall (QH) droplet} with area $\action=2\pi N\ell^2$ and uniform bulk density $\rho=1/2\pi\ell^2$. A similar picture is expected for fractional QH droplets confined by a weak potential \cite{Tsui,Laughlin:1983anomalous}, then with total area $\action=2\pi N\ell^2/\nu$ and bulk density $\rho=\nu/2\pi\ell^2$, both depending on the filling fraction $\nu$.

One of the hallmarks of topology in QH droplets is that they are bulk insulators but admit gapless edge modes, responsible for long-range correlations along the boundary. In particular, density-density edge correlations universally behave as
\beq
\frac{1}{\sin^2\bigl([\theta_1-\theta_2]/2\bigr)}
\qquad \text{(QH)}
\label{s15h}
\eeq
in terms of the canonical angle \eqref{sa} along the edge \cite{OLMSE:2024, Shabtai:2025}. (The actual correlator includes prefactors that are irrelevant here; \cf Eq.~\eqref{cedge} or the SM \cite{SM} for details.) Equivalently, low-energy edge modes propagate at a constant angular velocity in the $\theta$ coordinate \cite{OLMSE:2024}, similar to that of guiding centers \eqref{s15s}. Note that this is also expected to hold for fractional QH droplets, underscoring once more the significance of the area-preserving map \eqref{t15s}. The latter determines all the edge properties, hence the low-energy physics, of QH systems in the semiclassical limit.

\medskip
\noindent\textbf{Coulomb gases.} Consider now $N$ classical point charges in the plane with positions $\bx_1,\ldots,\bx_N$, interacting through a pairwise repulsive logarithmic potential, with no kinetic energy. The system's total energy is
\beq
E(\bx_1,\ldots,\bx_N)
= - \sum_{\!\!\!\!\!1\leq i<j\leq N\!\!\!\!\!}
    \log|\bx_i-\bx_j| + \frac{N}{2}\sum_{j=1}^N W(\bx_j),
\label{s1o}
\eeq
where $W$ is a confining potential that grows faster than $\log|\bx|$ as $|\bx|\to\infty$. When such a system is prepared at a finite temperature $\beta^{-1}$, it is known as a 2D \emph{Coulomb gas (CG)} or \emph{log-gas}. Its link with QH droplets stems from a coincidence, known as the \emph{plasma analogy}: when $W(\bx) = |\bx|^2$ is isotropic and harmonic, the probability density of a
configuration $\bx_1,\ldots,\bx_N$ at temperature $\beta^{-1}$ coincides with the $N$-body probability density of the Laughlin wave function at filling fraction $\nu = 2/\beta$ \cite{Laughlin:1983anomalous}. As we shall argue, the analogy generally fails to extend to anisotropic states, at least as far as edge properties are concerned.

In the thermodynamic limit $N\gg1$, the CG ground state is the configuration that minimizes Helmholtz free energy. In practice, energy dominates entropy due to the long-range interaction in Eq.~\eqref{s1o} \cite{Dyson:1962}, so it suffices to minimize energy alone, regardless of temperature. This is most easily described in terms of the (normalized) mean density $\rho(\bx) \equiv \frac{1}{N} \bigl\langle \sum_{j=1}^N \delta(\bx-\bx_j) \bigr\rangle$, where the expectation value $\langle \cdot \rangle$ is taken in the canonical ensemble. When $N\to\infty$, the density is supported in some domain $\droplet$ \footnote{We use the same letter $\droplet$ to denote both the domain occupied by a QH droplet and that occupied by a CG, even though they are \emph{a priori} unrelated. The plasma analogy ultimately requires choosing the domains to be identical.}. Minimizing the energy \eqref{s1o} over all density profiles yields the Euler-Lagrange equation \cite{SM}
\begin{align}
\label{e7}
\frac{W(\bx)}{2} - \int_{\droplet} \dd\by\, \rho(\by) \log |\bx-\by|
=
\text{const},
\end{align}
valid for all $\bx\in\droplet$, where $\dd\by\equiv\dd y^1\dd y^2$. Taking the Laplacian of both sides gives $\rho = \nabla^2 W / 4\pi$ inside the droplet and $\rho=0$ outside, which fixes the density. In turn, plugging $\rho = \nabla^2 W / 4\pi$ back into Eq.~\eqref{e7} implicitly determines $\droplet$ through a (generally complicated) inverse problem. The droplet's shape is thus determined by the potential $W$, but in contrast to QH droplets, it is also strongly affected by Coulomb interactions. (This is why our use of $V$ vs $W$ is important to distinguish QH and CG potentials.) A much easier task is to first choose $\droplet$ and a normalized $\rho$ therein, then reverse-engineer the potential $W$ in $\droplet$ through Eq.~\eqref{e7}. In fact, this is precisely how CGs with uniform bulk density are meant to be used in the plasma analogy.

Note that the left-hand side of Eq.~\eqref{e7} is the total electrostatic potential at $\bx$: it is the sum of the external potential $W$ and the mean potential $\braket{\Phi(\bx)}$ created by the CG itself, with
\begin{align}
\label{eq:potfield}
\Phi(\bx)
\equiv
-\sum_{j=1}^N \log |\bx-\bx_j|.
\end{align}
In that sense, Eq.~\eqref{e7} is a screening condition: the CG behaves as a perfect conductor. It implies that CGs are governed by 2D electrostatics in the thermodynamic limit.

\begin{figure}
\includegraphics[width=0.48\textwidth]{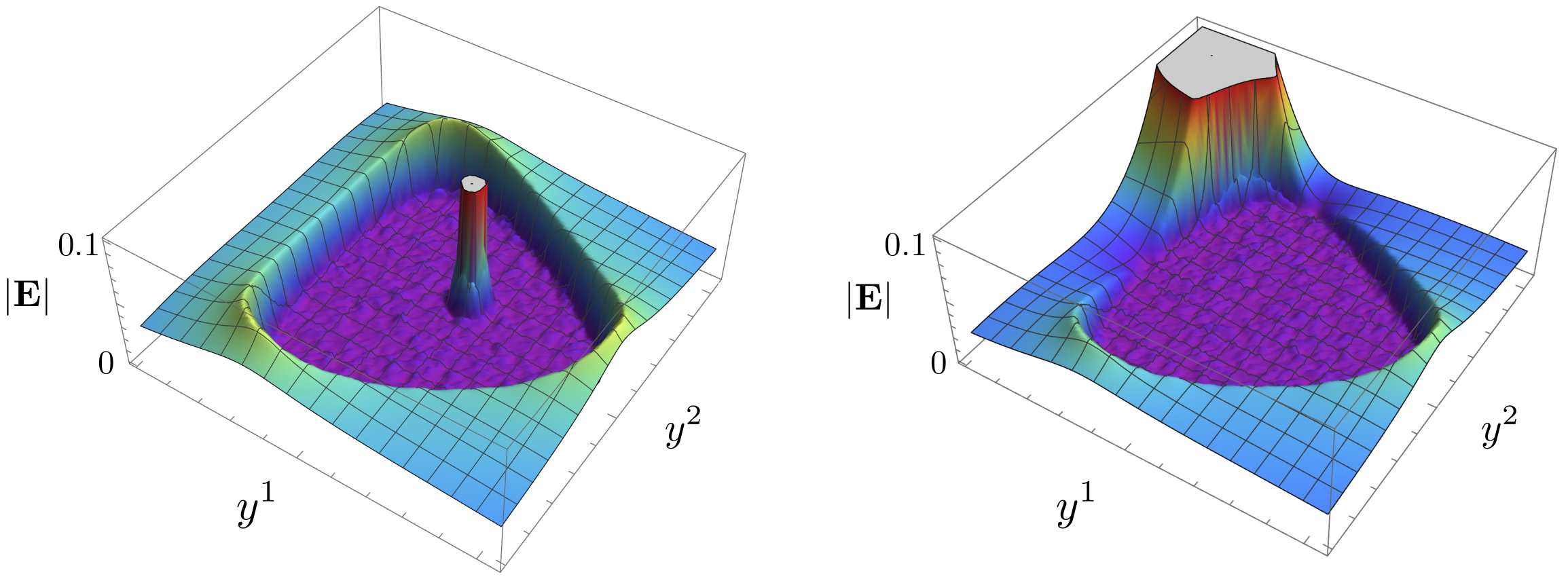}
\caption{Monte Carlo simulation of the norm of the total electric field $\mathbf{E}(\by) = -\nabla_{\by}\bigl(\braket{\Phi(\by)}^{\bx}-\braket{\Phi(\by)}-\log |\bx-\by|\bigr)$ for $N=256$ particles and potential $W(x,y) = x^2 + y^2 - (x^3-3xy^2)/10$, shown as a function of $\by =(y^1, y^2)$. (The potential is polynomial so as to give an anisotropic CG at minimal numerical cost, but it is otherwise unimportant.) Left: Fixed unit charge at $\bx\in\droplet$; its effect is screened over short distances. Right: Fixed unit charge at $\bx\notin\droplet$, in which case the electric potential is a long-range Green's function. In both plots, the electric field vanishes inside $\droplet$ and is discontinuous on $\partial \droplet$, signaling charge accumulation on the boundary.}
\label{fig:screening}
\end{figure}

A simple way to understand the relevance of conformal geometry outside CGs and on their edge is to study correlations. In that context, the fundamental object is the (connected) two-point function of the field in Eq.~\eqref{eq:potfield}, which can be determined from linear response. Indeed, suppose a (small) charge $q$ is inserted at some position $\bx$, so that the energy $E$ in Eq.~\eqref{s1o} is modified to $E+q\Phi(\bx)$. Denoting by $\braket{\cdot}^{\bx}$ the corresponding expectation value, linear response yields
\begin{align}
\label{e88}
\braket{\Phi(\by)}^{\bx}-\braket{\Phi(\by)}
\sim
-q \beta \braket{\Phi(\bx)\Phi(\by)}_{c}.
\end{align}
Here, ${\langle \cdot \rangle}_{c}$ denotes the connected correlator, and the left-hand side is the potential indirectly created by the charge $q$ at some other point $\by$, without including the potential $-q\log|\bx-\by|$ due to the point charge itself. By screening, the CG acts as a perfect conductor, so computing the left-hand side of Eq.~\eqref{e88} boils down to an electrostatic problem whose solution depends on whether $\bx,\by$ are inside or outside the droplet; see Fig.~\ref{fig:screening} and the SM \cite{SM} for details. When $\bx,\by$ are inside, the total potential $\braket{\Phi(\by)}^{\bx}-\braket{\Phi(\by)}-q\log |\bx-\by|$ is constant and its gradient, the total electric field, vanishes in $\droplet$ as in any conductor. When $\bx$ is inside and $\by$ outside (or vice versa), screening implies that the total field only depends on the point outside. However, when both $\bx$, $\by$ are outside the droplet, the total potential is the Green's function $\greensFunc(\bx,\by)$ of the Laplacian $-(1/q)\nabla^2$ outside $\droplet$, with Dirichlet boundary conditions on $\partial \droplet$.

If $\droplet$ is the unit disk, the corresponding Green's function $\greensFunc_0(\bx,\by)$ can be found explicitly through the method of images. Green's functions in more complicated geometries are then related to $\greensFunc_0$ by conformal maps \cite{SM}. Specifically, the relevant conformal transformation
\beq
\conformal:
\RR^2\setminus\droplet \to \RR^2\setminus\mathbb{D}:
\bx \mapsto\conformal(\bx)
\label{ss15n}
\eeq
maps the exterior of the droplet on the exterior of the disk \footnote{For comparison with QH droplets, we assume that the disk $\mathbb{D}$ and the droplet $\droplet$ in Eq.~\eqref{ss15n} have the same area. This entails no loss of generality, since it can be ensured by composing $\conformal(\bx)$ with an overall dilation that does not affect the angle $\alpha$ introduced below the map \eqref{ss15n}.}, whereupon the Green's function outside $\droplet$ is $\greensFunc(\bx,\by) = \greensFunc_0(\conformal(\bx),\conformal(\by))$. This provides a natural \emph{conformal angle coordinate}, which we denote $\alpha$: for $\bx \in \partial \droplet$, it is the polar angle of the vector $\conformal(\bx)$.

We stress that conformal maps such as in Eq.~\eqref{ss15n} are extremely rigid since they must be invertible and holomorphic (in the coordinate $x^1+\ii x^2$) everywhere outside $\droplet$. Indeed, provided the map \eqref{ss15n} sends infinity on itself, the Riemann mapping theorem ensures that it is unique up to an overall rotation. Put differently, all correlations in a CG are fully determined by the shape of the droplet $\droplet$.

An important application of the conformal map \eqref{ss15n} is provided by density-density correlations, especially since they can be compared with their counterparts in QH droplets. While bulk correlations are short-ranged due to screening \cite{Jancovici:1981}, edge correlations are long-ranged. They can be predicted from electrostatics: they are, in fact, correlations of the edge charge density produced at the boundary of the CG by the discontinuous electric field; see again Fig.~\ref{fig:screening}. The resulting correlator behaves as
\begin{align}
\label{s15n}
\frac{1}{\sin^2 \bigl( [\alpha_1-\alpha_2]/2 \bigr)}
\qquad \text{(CG)}
\end{align}
in terms of the conformal angle $\alpha$ along the edge, determined by the map \eqref{ss15n} \cite{Jancovici:1995, LebleSerfaty:2018, HedenmalmWennman:2021, AmeurCronvall:2023}. (As in Eq.~\eqref{s15h}, prefactors are omitted here; see the SM \cite{SM} for details.)

\medskip
\noindent\textbf{Mismatch.}
The difference between QH droplets and CGs can be summarized in terms of the maps in Eqs.~\eqref{t15s} and \eqref{ss15n}. Consider a (simply connected) domain $\droplet$ in the plane, and compare two scenarios: one where it is occupied by a QH droplet in a potential $V$, and another where it is occupied by a CG in a potential $W$. Both systems share the same boundary $\partial \droplet$. The density correlator between points $\bx_1$ and $\bx_2$ on $\partial \droplet$ is given by Eq.~\eqref{s15h} with $\theta_i = \theta(\bx_i)$ for QH, and by Eq.~\eqref{s15n} with $\alpha_i = \alpha(\bx_i)$ for CG. The coordinates $\theta_i$ and $\alpha_i$ are respectively derived from the area-preserving map \eqref{t15s} and the angle-preserving map \eqref{ss15n}. These maps usually differ, except in a few highly specific cases, causing the discrepancy between Eqs.~\eqref{s15h} and \eqref{s15n} despite their superficial similarity. Physically, this is because electrostatics is only sensitive to the shape of a CG, while QH physics is sensitive to both the shape and the local gradient of the confining potential at the edge.

This mismatch is our main statement, so let us rephrase it for later reference. Namely, the maps in Eqs.~\eqref{t15s} and \eqref{ss15n} yield two parameterizations of the edge:
\begin{align}
\label{s1da}
\theta:
\partial \droplet \to S^1:
\bx \mapsto \theta(\bx)
& \quad \text{(QH)},\\
\label{s1db}
\alpha: 
\partial \droplet \to S^1:
\bx \mapsto \alpha(\bx)
& \quad \text{(CG)}.
\end{align}
These typically differ, in the sense that the one-dimensional diffeomorphism $\alpha\circ\theta^{-1}: S^1 \to S^1$ is not just a rotation $\varphi \mapsto \varphi + \textrm{const}$ (except in rare special cases). In the remainder of this paper, we verify this claim in several situations of interest and point out its experimental ramifications. We stress that this does not affect the universal correspondence between bulk topology and edge modes \cite{Hatsugai:1993}: indeed, as shown in \cite{OLMSE:2024}, the emergent effective edge theory is unaffected by our deformations of the sample's geometry (at least to leading order in $N$).

\section{\texorpdfstring{Mismatched edge correlations\\in anisotropic droplets}{Mismatched edge correlations in anisotropic droplets}}
\label{se3}

Below, we provide detailed examples of the difference between the angle coordinates in Eqs.~\eqref{s1da} and \eqref{s1db}. We begin with disk-shaped QH droplets whose confining potential is anisotropic, then consider anisotropic droplets of the form introduced in \cite{OLMSE:2024}, including square droplets where CG predictions are known analytically.

\begin{figure}[t]
\includegraphics[width=0.48\textwidth]{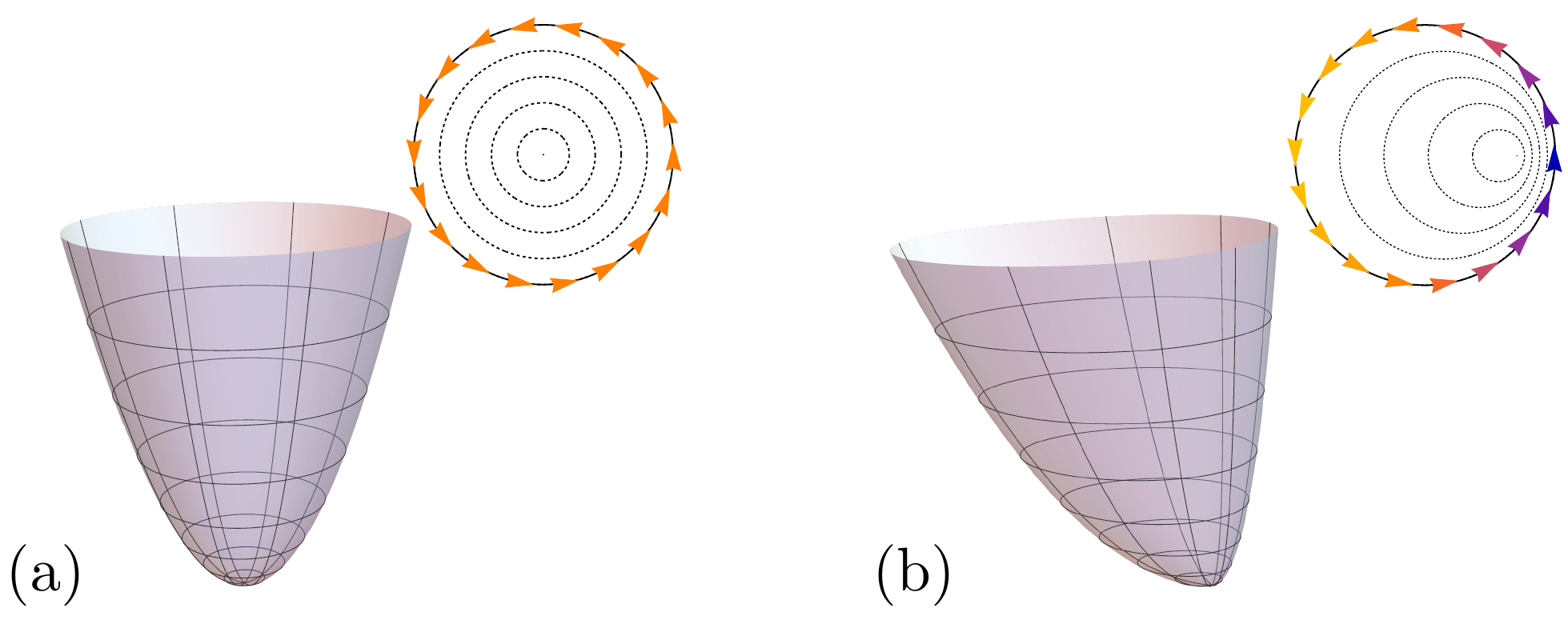}
\caption{Two confining potentials, with their level curves as insets. The one in (a) is isotropic, so the norm \eqref{v_theta} of the edge velocity is uniform in the resulting QH droplet. The one in (b) is anisotropic, but at least one of its equipotentials is a circle, so it gives rise to a QH droplet whose edge modes have an inhomogeneous velocity \eqref{v_theta} despite the droplet's isotropic shape. QH physics distinguishes these two situations. By contrast, from the CG perspective, the two systems are equivalent since edge properties are determined by the droplet's shape alone.}
\label{fig2}
\end{figure}

\medskip
\noindent\textbf{Anisotropic disks.}
Consider a QH droplet whose domain $\droplet$ is a disk but whose confining potential is anisotropic as in Fig.~\ref{fig2}(b) (so its level curves are not concentric circles). Edge modes then propagate with a velocity whose norm \eqref{v_theta} is nonconstant, leading to a canonical angle \eqref{sa}. When the droplet $\droplet$ is a disk, the arc length is $s=L\varphi/2\pi$, and the angle \eqref{sa} becomes a nontrivial function $\theta(\varphi)$, expressed as a functional of the local norm $v(\varphi)$ of the edge velocity. The resulting map \eqref{s1da} is a nontrivial diffeomorphism of the circle since $\partial \droplet = S^1$.

Now suppose one were to use the plasma analogy to predict edge correlations in this setup. Consider therefore a disk-shaped CG with uniform bulk density. The corresponding potential in the screening condition \eqref{e7} is necessarily isotropic, and the conformal map \eqref{ss15n} is just the identity (up to rotation), so the coordinate $\alpha$ in the edge correlator \eqref{s15n} coincides with the polar angle: $\alpha=\varphi$. This manifestly differs from the canonical angle \eqref{sa}, exemplifying the announced mismatch.

Note that potentials giving rise to disk-shaped droplets with anisotropic edge velocities can be built explicitly. Indeed, let $\theta(\varphi)$ be any smooth function such that \footnote{Geometrically, any smooth real function that satisfies Eq.~\eqref{tc} defines an (orientation-preserving) diffeomorphism of the circle.}
\beq
\theta(\varphi+2\pi)=\theta(\varphi)+2\pi,
\qquad
\theta'(\varphi)>0.
\label{tc}
\eeq
Then, choose a radius $R$ for the QH droplet, and let the potential near the edge take the form
\beq
V(r,\varphi)
=
\Vnull\bigl(\pi R^2 + \pi \bigl[r^2-R^2\bigr] \big/ \theta'(\varphi)\bigr)
\label{sc}
\eeq
for some monotonically increasing function $\Vnull(\cdot)$. The corresponding action-angle coordinates in a neighborhood of the edge are
\beq
\action
=
\pi R^2 + \pi \bigl[r^2-R^2\bigr] \big/ \theta'(\varphi),
\qquad
\theta = \theta(\varphi).
\label{ssc}
\eeq
Fixing the Fermi energy at $\Vnull(\pi R^2)$, the edge of the droplet is the circle $r=R$, but the norm \eqref{v_theta} of the edge velocity is generally anisotropic since $v(\varphi)=\text{const}/\theta'(\varphi)$.

\medskip
\noindent\textbf{Flower droplets.} Let us turn to anisotropic QH droplets confined by \emph{edge-deformed potentials} as in \cite{OLMSE:2024}: given any smooth function $\theta(\varphi)$ satisfying Eq.~\eqref{tc}, define
\beq
V(r,\varphi)
\equiv
\Vnull\bigl(\pi r^2 / \theta'(\varphi)\bigr),
\label{s1e}
\eeq
where $\Vnull(\cdot)$ is again a monotonically increasing function. The corresponding action-angle coordinates are given by Eq.~\eqref{ssc} with $R=0$, and the equipotentials
\beq
\label{eq:edge}
r^2
= \frac{\action}{\pi} \, \theta'(\varphi)
\eeq
are nested as in Fig.~\ref{fig1}. An advantage of this family of potentials is that their semiclassical eigenstates in the LLL can be found analytically, for any $\theta(\varphi)$ \cite{OLMSE:2024}.

\begin{figure}[t]
\includegraphics[width=0.47\textwidth]{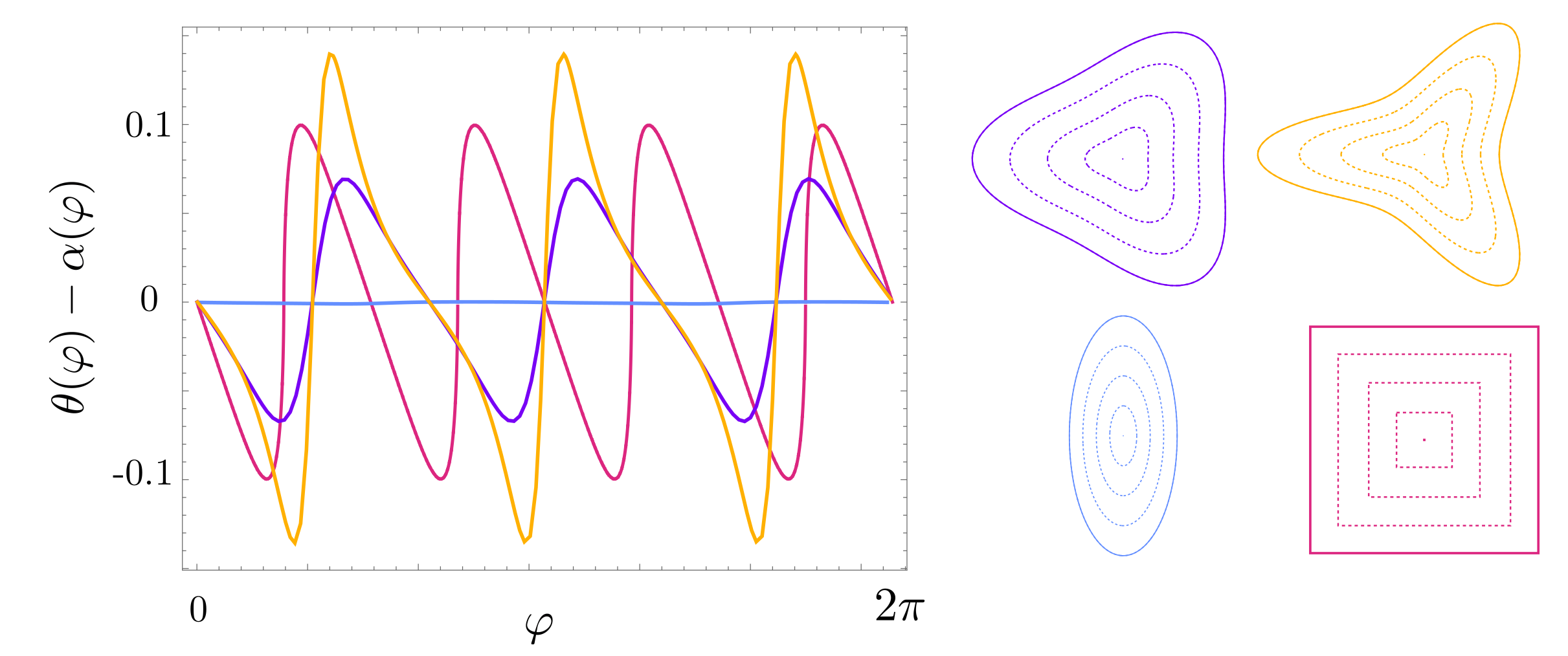}
\caption{Differences between coordinates $\theta(\varphi)$ and $\alpha(\varphi)$ for edge-deformed droplets \eqref{eq:edge} given by flower deformations \eqref{s1z}, with $(k, \lambda) = (3, 0.2)$ (purple), $(k, \lambda) = (3, 0.4)$ (yellow), $(k, \lambda) = (2, 0.4)$ (blue), and a square \eqref{s15y} (magenta). Equipotentials of the respective QH droplets are shown on the side.}
\label{fig3}
\end{figure}

A subclass of edge-deformed QH droplets is provided by \emph{flower deformations} with an integer number $k$ of petals, for which the function $\theta(\varphi)$ in the potential \eqref{s1e} is given by
\beq
\ee^{\ii k\theta(\varphi)}
=
\frac{\cosh(\lambda) \ee^{\ii k\varphi}+\sinh(\lambda)}{\sinh(\lambda) \ee^{\ii k\varphi}+\cosh(\lambda)},
\label{s1z}
\eeq
where $\lambda\in\RR$ is a deformation parameter. (The cases $k=2$ and $3$ are shown in the inset of Fig.~\ref{fig3}.) It is then straightforward to express the QH density-density correlator \eqref{s15h} in terms of the polar angle $\varphi$, for any $k$ and $\lambda$. For $k=2$ and $\Vnull(\action)\propto\action$, the potential \eqref{s1e} is actually harmonic [and anisotropic if $\lambda\neq0$ in Eq.~\eqref{s1z}].

Now suppose again one uses the plasma analogy to describe this situation. To this end, consider a CG with uniform bulk density, whose shape $\droplet$ is a flower droplet with a boundary given by Eq.~\eqref{eq:edge} in polar coordinates, with some fixed area $\action$. The resulting edge density correlator takes the form \eqref{s15n} in terms of the restriction of the conformal map \eqref{ss15n} to the boundary $\partial\droplet$. The problem is to find this map $\bx\mapsto\conformal(\bx)$. As before, $\conformal(\bx)$ is uniquely fixed by the shape of the droplet through the Riemann mapping theorem. It admits no known analytical expression (unless $k = 2$), but it can be built numerically using Fekete points \cite{SaffTotik:1997}. This is how Fig.~\ref{fig3} was obtained: in it, we show the difference $\theta(\varphi)-\alpha(\varphi)$ for flower droplets with $k = 2$ or $3$ petals. The difference is manifestly nonconstant when $k=3$, confirming that $\theta(\varphi)$ and $\alpha(\varphi)$ differ by more than a rotation. In fact, the mismatch worsens when $\lambda$ increases. Similar conclusions hold for all droplets with $k\neq2$ petals and generally for the vast majority of QH droplets confined by edge-deformed potentials \eqref{s1e}.

The exception $k=2$ corresponds to droplets whose boundary, given by Eq.~\eqref{eq:edge}, is an ellipse. In that case, the conformal map \eqref{ss15n} can be found analytically \cite{SM}, and the CG edge correlator \eqref{s15n} coincides with the correlator \eqref{s15h} of a QH droplet in the potential \eqref{s1e}; see the straight line where $\theta(\varphi)=\alpha(\varphi)$ in Fig.~\ref{fig3}. There is a reason for this coincidence: the distribution of particles in an elliptic CG coincides with the distribution of eigenvalues of random matrices in the elliptic Ginibre ensemble \cite{Girko:1986, ForresterJancovici:1996}, which in turn coincides with the density of LLL wave functions in a harmonic potential.

A geometric way to understand why $k=2$ is exceptional is to bluntly impose $\theta(\varphi)=\alpha(\varphi)$ on the edge given by Eq.~\eqref{eq:edge}, then attempt to view $\alpha(\varphi)$ as the boundary value of a bijective conformal map \eqref{ss15n}. One readily verifies that this fails for any flower deformation \eqref{s1z}, unless $k=2$. The failure is due to zeros or poles of the would-be conformal extension, which always appear outside $\droplet$ unless $k=2$; see the SM \cite{SM} for details. More generally, for any edge-deformed droplet, imposing $\theta(\varphi)=\alpha(\varphi)$ on the curve \eqref{eq:edge} makes it impossible to view $\alpha(\varphi)$ as the restriction of a bijective conformal map \eqref{ss15n}, except in exceedingly fine-tuned cases. This underscores again that the plasma analogy is typically unreliable on the edge of anisotropic QH droplets.

\medskip
\noindent\textbf{Square droplets.}
We finally turn to a case where both QH and CG can be treated analytically, namely a square droplet. On the QH side, we build the droplet by choosing an edge-deformed potential \eqref{s1e} whose equipotentials are squares. The corresponding canonical angle coordinate is
\begin{align}
\label{s15y}
\theta(\varphi)
=
q(\varphi)+\frac{\pi}{4}\tan\bigl(\varphi-q(\varphi)\bigr),
\quad
q(\varphi)
\equiv
\frac{\pi}{2}
\left\lfloor\frac{2\varphi}{\pi}+\frac{1}{2}\right\rfloor,
\end{align}
using the floor function $\lfloor\cdot\rfloor$. On the CG side, the problem is to find the conformal map \eqref{ss15n} that sends the outside of a square on the outside of a disk, thereby fixing the CG angle \eqref{s1db}. The solution is well known: it is given by a Schwarz-Christoffel map \cite{DriscollTrefethen:2002} whose explicit form involves a hypergeometric function \cite{SM}. The ensuing CG angle can be expressed as
\beq
\label{alpha_square}
\alpha(\varphi)
=
-\ii\log\left[\conformal\left(\frac{\sqrt{\pi}}{2} \frac{\ee^{\ii\varphi}}{|\cos (\varphi+q(\varphi))|}\right)\right]
\eeq
in terms of the polar angle $\varphi$ and the map $\conformal$ in Eq.~\eqref{ss15n}, which is holomorphic in the complex coordinate $x + \ii y$. One can again compare the QH and CG angles in Eqs.~\eqref{s15y} and \eqref{alpha_square}: their difference $\theta(\varphi)-\alpha(\varphi)$, shown in Fig.~\ref{fig3}, fails to be a constant, exhibiting once more the QH/CG mismatch.

\section{\texorpdfstring{Consequences for\\fluctuations and admittance}{Consequences for fluctuations and admittance}}
\label{se5}

Differences between QH droplets and their would-be plasma analogues affect the computation of physical quantities. Indeed, the mismatch between the correlators in Eqs.~\eqref{s15h} and \eqref{s15n} entails different predictions for the contribution of low-energy edge modes to, say, linear-response coefficients of anisotropic droplets. Here, we demonstrate its consequences for fluctuations of local operators, known as \emph{linear statistics} in the CG literature \cite{ZabrodinWiegmann:2006, LebleSerfaty:2018, DeBruyneEtAl:2024}. As we shall explain, the fluctuation-dissipation theorem makes it possible to probe such fluctuations through microwave absorption experiments that are within reach using current technologies \cite{Talyanskii, Andreev, Cano, Mahoney, FrigerioEtAl:2024}.

\medskip
\noindent\textbf{Fluctuations in quantum Hall droplets.} To begin, consider the QH side. Given a smooth real function $\LSFunc(\bx)$, define the Fock-space operator
\beq
\LSOp
\equiv
\int \dd\bx\, \LSFunc(\bx) \hat c^{\dagger}(\bx)\hat c(\bx),
\label{s1x}
\eeq
where $\hat c^{(\dagger)}(\bx)$ are local fermionic annihilation (creation) operators satisfying $\{\hat c(\bx_1),\hat c^{\dagger}(\bx_2)\}=\delta^{(2)}(\bx_1-\bx_2)$. We wish to evaluate the variance of $\LSOp$ in the $\nu = 1$ QH ground state. By Wick's theorem, one can express the variance as
\beq
\Var[\LSOp]_{\textrm{QH}}
=
\frac{1}{2} \iint \dd\bx\,\dd\by\, 
|C(\bx,\by)|^2 \bigl[\LSFunc(\bx)-\LSFunc(\by)\bigr]^2,
\label{s15x}
\eeq
where $C(\bx,\by)$ is the electronic two-point correlation function. The latter is short-ranged in the bulk, where
\beq
|C(\bx,\by)|_{\textrm{bulk}}
\sim
\frac{1}{2\pi\ell^2} \exp \biggl( -\frac{|\bx-\by|^2}{4\ell^2} \biggr)
\label{t15x}
\eeq
with $\ell$ the magnetic length introduced below Eq.~\eqref{t15s}. By contrast, correlations are long-ranged along the edge, as mentioned around Eq.~\eqref{s15h}. A more precise statement, derived in \cite{OLMSE:2024} for edge-deformed potentials, is
\beq
\label{cedge}
|C(\bx,\by)|_{\textrm{edge}}
\sim
\frac{\mathcal{N}}{\sqrt{v(\theta_{\bx}) v(\theta_{\by})}}
\frac{\exp \Bigl( -\frac{d_{\bx}^2}{2\ell^2}-\frac{d_{\by}^2}{2\ell^2} \Bigr)}{2\sin\bigl(|\theta_{\bx}-\theta_{\by}|/2\bigr)}
\eeq
up to an omitted normalization $\mathcal{N}$ \cite{SM}, where $\theta_{\bx}$ is the canonical angle coordinate \eqref{s1da} of $\bx$, $d_{\bx}$ is the Euclidean distance between $\bx$ and the edge, $v(\theta_{\bx})$ is the norm \eqref{v_theta} of the edge velocity at $\theta_{\bx}$, and similarly for $\by$. Plugging this and Eq.~\eqref{t15x} into the variance \eqref{s15x}, and assuming that the function $\LSFunc(\bx)$ varies slowly on length scales of the order of $\ell$,
yields
\beq
\label{s1w}
\Var[\LSOp]_{\textrm{QH}}
\sim
\int_\droplet \frac{\dd\bx}{4\pi}\,|\nabla \LSFunc|^2
+ \oint\frac{\dd\theta_1\dd\theta_2}{32\pi^2}
  \frac{\bigl(\LSFunc_{\theta_1}-\LSFunc_{\theta_2}\bigr)^2}{\sin^2\biggl(\displaystyle \frac{\theta_1-\theta_2}{2}\biggr)},
\eeq
where $\LSFunc_{\theta}$ is the function of $\theta$ obtained by restricting $\LSFunc(\bx)$ to the edge and labeling points by their canonical angle. The result \eqref{s1w} holds at leading order in the thermodynamic limit $N\to\infty$ with $N\ell^2$ kept finite. It exhibits a splitting between bulk and edge contributions, the latter being sensitive to the norm \eqref{v_theta} of the local edge velocity through the coordinate $\theta$.

\medskip
\noindent\textbf{Fluctuations in Coulomb gases.}
On the CG side, the plasma analogue of the quantum variance \eqref{s15x} is the thermal fluctuation of the linear statistics $\sum_{j=1}^{N} \LSFunc(\bx_j)$, where $\LSFunc$ is the same function as in Eq.~\eqref{s1x}. One is usually interested in the thermodynamic limit,
\beq
\Var[\LSFunc]_{\textrm{CG}}
\equiv
\lim_{N\to\infty}
\Var\!\left[\sum_{j=1}^{N} \LSFunc(\bx_j)\right]\!.
\label{s15w}
\eeq
The latter can be split into bulk and edge correlations in direct analogy to Eq.~\eqref{s1w}. Indeed, it was shown in \cite{ZabrodinWiegmann:2006, AmeurHedenmalmMakarov:2015, LebleSerfaty:2018} that the CG variance \eqref{s15w} is \cite{SM}
\beq
\Var[\LSFunc]_{\textrm{CG}}
= \int_{\droplet}\frac{\dd\bx}{2\pi\beta}\,|\nabla \LSFunc|^2
+ \oint\frac{\dd\alpha_1\dd\alpha_2}{16\pi^2\beta}
  \frac{
     \bigl(\LSFunc_{\alpha_1}-\LSFunc_{\alpha_2}\bigr)^2
  }{
    \sin^2\biggl(\displaystyle \frac{\alpha_1-\alpha_2}{2}\biggr)
  },
\label{t15w}
\eeq
where $\LSFunc_{\alpha}$ is defined analogously to $\LSFunc_{\theta}$ in Eq.~\eqref{s1w}, but using the conformal coordinate \eqref{s1db} instead of the canonical one \eqref{s1da}. For the free-fermion value $\beta=2$, Eq.~\eqref{t15w} looks identical to Eq.~\eqref{s1w}, up to one key difference: the angles $\theta,\alpha$ stem from different maps---\eqref{t15s} vs \eqref{ss15n}---and thus lead to distinct functions $\LSFunc_{\theta}$, $\LSFunc_{\alpha}$.

\begin{figure}[htbp]
\includegraphics[width=0.483\textwidth]{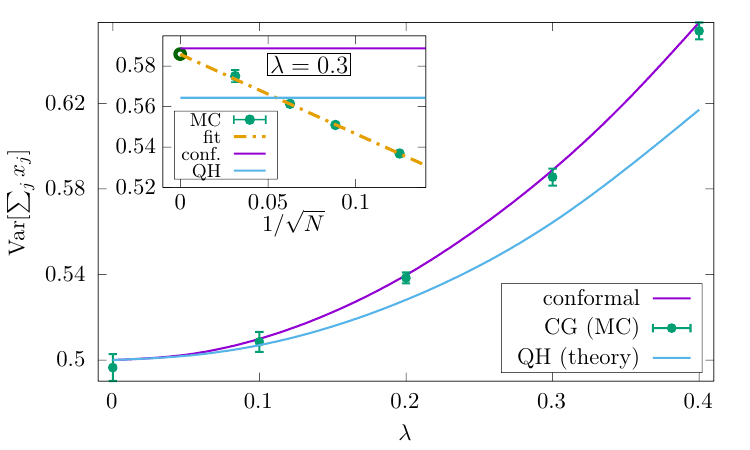}
\caption{The total variance of $\LSFunc(x, y) = x$ for flower droplets with $k=3$ petals and $\lambda$ ranging from $0$ to $0.4$. Main figure: The variance \eqref{s1w} for QH droplets is compared with the CG result \eqref{t15w} obtained from a numerical evaluation of the conformal map \eqref{ss15n}, including an extrapolation based on Monte Carlo (MC) data for the CG at $\lambda = 0$, $0.1$, $0.2$, $0.3$, and $0.4$ \cite{SM}. The two match for $\lambda=0$ but not $\lambda>0$. Inset: Details of the extrapolation for $\lambda=0.3$. The variance is extrapolated by a fit of the form $a+b/\sqrt{N}$ for $N = 64$, $128$, and $256$ particles. The intercept with the $y$ axis yields the coefficient $a$, which is shown as an empty green circle and differs significantly from its QH counterpart. To further illustrate this, we include an MC point for $N=1024$, which has a larger error bar but remains consistent with a leading $1/\sqrt{N}$ correction.}
\label{fig5}
\end{figure}

\medskip
\noindent\textbf{Mismatch.}
Returning to the examples of Sec.~\ref{se3}, it is straightforward to illustrate the quantitative mismatch between the fluctuations in Eqs.~\eqref{s1w} and \eqref{t15w}. Indeed, take for simplicity the function $\LSFunc(x,y) = x$ for which $|\nabla \LSFunc|^2 = 1$, so that the bulk terms in Eqs.~\eqref{s1w} and \eqref{t15w} are proportional to the droplet's area. The resulting total variances are plotted in Fig.~\ref{fig5} for flowers with $k=3$ petals in Eqs.~\eqref{eq:edge}--\eqref{s1z}, showing that QH and CG predictions disagree and that the mismatch worsens as the anisotropy increases. The same conclusions can be drawn for the square droplet: in that case, the variances in Eqs.~\eqref{s1w} and \eqref{t15w} turn out to be
\beq
\label{VarU_QHCG}
\Var[\LSOp]_{\textrm{QH}}
\approx
0.5183083,
\qquad
\Var[\LSFunc]_{\textrm{CG}}
\approx
0.5469944,
\eeq
which manifestly differ. Their exact values can be expressed using an infinite series and a generalized hypergeometric function, respectively \cite{SM}.

\medskip
\noindent\textbf{Microwave absorption.}
One way to measure the edge contribution to the QH fluctuation \eqref{s1w} is through microwave absorption experiments \cite{Talyanskii, Andreev, Cano, Mahoney, FrigerioEtAl:2024}. Indeed, consider a QH droplet perturbed by an electrostatic potential energy $\LSFunc(\bx) \cos(\omega t)$, with driving frequency $\omega>0$ close to the angular Fermi velocity $\omega_{\textrm{F}}$ of edge modes \footnote{The angular Fermi velocity $\omega_{\textrm{F}}$ typically lies in the microwave range in experiments where this has been carried out; see \eg \cite{KamataEtAl:2010}.}. Time-dependent perturbation theory applied to low-energy eigenstates then predicts that the droplet absorbs this radiation at a rate \cite{OLMSE:2024}
\beq
\Gamma(\omega)
\sim
\frac{1}{2\hbar^2}
\! \sum_{p=1}^{\infty} p |\LSFunc_p|^2 \delta(\omega-p\hspace{1pt}\omega_{\textrm{F}}),
\label{s15v}
\eeq
where $\LSFunc_p \equiv \oint\frac{\dd\theta}{2\pi}\ee^{\ii p\theta} \LSFunc_{\theta}$ is the $p^{\text{th}}$ Fourier mode of the function $\LSFunc_{\theta}$ in Eq.~\eqref{s1w}. [Just as Eq.~\eqref{s1w}, the result \eqref{s15v} relies on the assumption that $\LSFunc(\bx)$ is almost constant on length scales of order $\ell$.] Integrating $\Gamma(\omega)$ over all frequencies yields
\beq
\label{eq:fludiss}
\int\limits_0^{\infty}\!\dd\omega\,\Gamma(\omega)
=
\frac{1}{2\hbar^2}
\! \sum_{p=1}^{\infty} p |\LSFunc_p|^2
=
\oint\frac{\dd\theta_1\dd\theta_2}{2\hbar^2}
\frac{
  \bigl( \LSFunc_{\theta_1}-\LSFunc_{\theta_2} \bigr)^2
}{
  \sin^2\biggl(\displaystyle \frac{\theta_1-\theta_2}{2}\biggr)
},
\eeq
which is nothing but the edge contribution to the variance \eqref{s1w}. In this sense, the latter is measurable in QH experiments, allowing direct verification of the deviation between QH and CG physics.

Note that the link in Eq.~\eqref{eq:fludiss} between absorption and variance was to be expected, given the fluctuation-dissipation theorem. In fact, the latter ensures that the full variance \eqref{s1w}---including the bulk term---is the integral of the full absorption rate, including effects of higher Landau levels. No such bulk term appears in Eq.~\eqref{eq:fludiss} since the rate \eqref{s15v} is limited to low-energy transitions within the LLL.

\section{Conclusions}
\label{se6}

In this paper, the plasma analogy was tested on integer QH droplets formed by anisotropic confining potentials. While it reproduces QH data in the bulk, the analogy generally fails at the edge---apart from exceptional, fine-tuned scenarios. Such discrepancies are expected to be directly relevant for experiments, both because of advances in measurement techniques for solid-state systems \cite{Hashimoto, Hashimoto2008, Mahoney, FrigerioEtAl:2024} and increases in the precision of quantum simulators \cite{FletcherEtAl:2021, Mukherjee:2021jjl, YaoEtAl:2024, Crepel:2023kbu, Leonard:2023ndq, Binanti:2023ozm, Braun:2024yjh}. In each case, deviations from leading-order results are becoming visible, which is where the conclusions of this paper are expected to be most striking.

We showed that the edge discrepancies stem from an elementary mismatch between QH and CG physics: the former is incompressible hydrodynamics, the latter is electrostatics. In geometric terms, QH is symplectic, while CG is conformal. This distinction entails different angular dependencies of the corresponding correlators in Eqs.~\eqref{s15h} and \eqref{s15n}, which in turn affects \eg fluctuations of local observables---quantities proposed to be measurable in the near future \cite{Talyanskii, Andreev, Cano, Mahoney, FrigerioEtAl:2024}. Such a breakdown of the plasma analogy is similarly expected near the edge of fractional QH states, as electron interactions are unlikely to improve the situation. It would be interesting to study this question for fractional droplets and other related systems where interactions play an important role, such as quantum dots or artificial atoms \cite{PhysRevLett.65.108, PhysRevB.66.041304, LiebSolovejYngvason:2001} and QH droplets with quasihole excitations \cite{LambertLundholmRougerie:2023}.

\section*{Acknowledgments}

We are grateful to Alexander Abanov, Yacin Ameur, Mathieu Beauvillain, Gabriel Cardoso, Laurent Charles, Benoit Dou\c{c}ot, Nathan Goldman, Gian Michele Graf, Hans Hansson, Leonardo Mazza, Gaultier Lambert, Alberto Nardin, Marios Petropoulos, Nicolas Rougerie, Gregory Schehr, Ood Shabtai, and Paul Wiegmann for fruitful discussions on closely related subjects. P.M.\ acknowledges financial support from the Wenner-Gren Foundations (Grant No.\ FT2022-0002). B.L.\ acknowledges financial support from the Swiss National Science Foundation (Postdoc.Mobility Grant No.\ 214461).


%


\onecolumngrid
\clearpage


\begin{center}
{\Large Supplemental Material}
\end{center}
\csname phantomsection\endcsname%
\addcontentsline{toc}{part}{Supplemental Material}%


This supplemental material has three parts. Parts~A and B respectively review the physics of quantum Hall (QH) droplets and Coulomb gases (CGs). Part~C complements the main text by showing that the parameterization of the edge by a canonical angle coordinate generally fails to extend to the biholomorphism required in the CG setting, despite the fact that local conformal maps do reproduce the correlations of edge-deformed QH droplets.

\section*{Part~A: Quantum Hall droplets}

Here, we review certain aspects of (anisotropic) QH droplets used in the main text. The presentation is limited to droplets of free fermions, although a similar picture is expected for fractional QH states as well. The key formulas are Eqs.~\eqref{eq:C_edge}--\eqref{e40b}, which are repeatedly used in the main text. We refer to \cite{OLMSE:2024} for more details.

\subsection*{Landau Hamiltonian with a trap}

Our starting point is the quantum mechanics of a charged particle (charge $q$, mass $M$) in two dimensions (2D). The particle is confined by some potential $V$ and placed in a perpendicular, uniform magnetic field $\bB = B\dd x\wedge\dd y$ (viewed for convenience as a two-form on the plane). The Hamiltonian of the particle is
\beq
\label{ham}
H
=
\frac{1}{2M}(\bp-q\bA)^2 + V(\bx),
\eeq
where $\bx$ denotes position, $\bp$ is canonical momentum, and $\bA$ is a vector potential such that $\dd\bA = \bB$ (viewing $\bA$ as a one-form). We work in the regime of strong magnetic fields, in the sense that $V(\bx)$ varies slowly on length scales of the order of the magnetic length
\beq
\ell \equiv \sqrt{\hbar/qB}
\label{eldef}
\eeq
(assuming $qB>0$ for definiteness). The problem is to find eigenstates of the Hamiltonian \eqref{ham}, and use them to build QH droplets consisting of a large number $N\gg1$ of free, spin-polarized electrons.

For vanishing potential $V=0$, the Hamiltonian \eqref{ham} is the usual Landau Hamiltonian of a free particle in a magnetic field. The energy spectrum then consists of discrete Landau levels, each of which is infinitely degenerate with energy $E_n=(n+1/2)\hbar qB/M$ in terms of a level index $n=0,1,2,\ldots$. The lowest Landau level (LLL) is the level $n=0$, separated from the level $n=1$ by the cyclotron gap $\hbar qB/M$. In symmetric gauge where $\bA = (Br^2/2)\dd\varphi$, an orthonormal basis of the LLL is provided by wave functions
\beq
\phi_m(\bx)
=
\frac{1}{\sqrt{2\pi\ell^2}}\frac{z^m}{\sqrt{m!}} \ee^{-|z|^2/2}
\eeq
labeled by an angular momentum quantum number $m=0,1,2,\ldots$, where $z\equiv(x + \ii y)/\sqrt{2}\ell$ is a dimensionless complex coordinate. The wave function of any state in the LLL is thus a holomorphic function times the Gaussian factor $\ee^{-|z|^2/2}$. For weak harmonic potentials, the story dates back to \cite{Fock:1928, Darwin:1931} and is closely related to the case $V = 0$; see \eg \cite{OLMSE:2024}.

We denote by $P=\sum_{m=0}^{\infty}|\phi_m\rangle\langle\phi_m|$ the LLL projection operator. One of its key properties is to make space noncommutative: $[PxP,PyP] = \ii \ell^2$. Thus, in the LLL, space becomes akin to a symplectic manifold and the magnetic length \eqref{eldef} is analogous to Planck's constant. We therefore refer to the limit of strong magnetic fields ($\ell\to0$) as a \emph{classical limit}.

\subsection*{Semiclassical wave functions}

In the limit of strong magnetic fields, the cyclotron gap diverges and the low-energy eigenstates of the full Hamiltonian \eqref{ham} belong to the LLL. It is thus meaningful to seek the spectrum of the LLL-projected Hamiltonian $PHP$, as opposed to the unprojected $H$. The projection $P$ essentially quashes the kinetic energy to a common zero-point value, so that the projected Hamiltonian boils down to a projected potential: $PHP=PVP+\text{const}$. It readily follows that semiclassical eigenstates of $H$ are localized on level curves of $V$. The issue is to understand which equipotentials are ``allowed'' by quantization, how localized the eigenstates are, and how their local norm and phase vary along the equipotential.

These questions were addressed in the mathematics literature \cite{Charles:2003a, Charles:2003b}, but they only recently gained appreciation in physics. Indeed, it was shown in \cite{OLMSE:2024} that semiclassical eigenstates can be built thanks to a holomorphic WKB ansatz that produces explicit formulas for wave functions, with many-body consequences that can be computed in closed form.

The results of \cite{OLMSE:2024, Charles:2003a, Charles:2003b} can be summarized as follows. Assume that the potential $V$ in Eq.~\eqref{ham} is \emph{monotonic} in the sense that it has a unique global minimum with nested level curves away from it. As explained in the main text, this implies the existence of global action-angle coordinates $(\action,\theta)$ in the plane, using which the potential can be written as $V(\bx) = \Vnull(\action(\bx))$. Eigenstates of the $PVP$ operator can then be labeled by the quantized area that they enclose, namely $\action_m\equiv2\pi m\ell^2$ with $m$ a large integer. (Technically, the semiclassical limit is such that $\ell\to0$ and $m\to\infty$ with $m\ell^2$ kept finite.) This label $m$ specifies the energy eigenvalue
\beq
\label{E_m}
E_{m}
=
\Vnull(2\pi m\ell^2)+O(\ell^2)
\eeq
and may be seen as generalizing angular momentum. Up to subleading corrections, the corresponding eigenstate of $PVP$ can be expressed as
\beq
\label{psi_m}
\psi_m(\bx)
\sim
\frac{\ee^{\ii m\theta+\ii\Theta_m(\bx)}}{\sqrt{2\pi^{3/2}\ell}}
\sqrt{\frac{\dot{\theta}(\action_{m})}{v_{m}(\theta)}}\,
\ee^{-d^2 / 2\ell^2}.
\eeq
This is written here in terms of the canonical angle $\theta$ on the equipotential $\action_{m} = 2\pi m\ell^2$, the distance $d$ from $\bx$ to the equipotential, the constant angular velocity $\dot{\theta}(\action_{m}) = \tfrac{2\pi}{qB}\Vnull'(\action_m)$ as in Eq.~\eqref{s15s}, and the Euclidean norm \eqref{v_theta} of the guiding-center drift velocity on the equipotential. The classical phase $m\theta$ gives all the winding of $\psi_m$ with $\theta$, and is accompanied by a slowly varying quantum correction $\Theta_m(\bx)$ that we leave unspecified here. Note the intuition in Eq.~\eqref{psi_m}: it confirms that the probability density $|\psi_m(\bx)|^2$ is localized as a Gaussian with width $\ell$ on the $m^{\text{th}}$ equipotential of $V$ and that the corresponding probability density $|\psi_m(\bx)|^2$ is proportional to the inverse of the guiding-center velocity. This agrees with expectations based on classical reasoning: the link between quantum wave function and classical local velocity is one of the hallmarks of the WKB method.

\subsection*{Correlations in anisotropic quantum Hall droplets}

Starting from the wave functions \eqref{psi_m}, it is straightforward to build a QH ground states of $N\gg1$ free electrons, since it is just a Slater determinant of occupied states $\psi_0,\psi_1,\ldots,\psi_{N-1}$ with $N\gg1$. (Strictly speaking, we have no access to wave functions $\psi_m$ for low quantum numbers $m$, but this is not an issue for edge properties, which occur at $m \sim N \gg 1$.) Since each wave function is localized on an equipotential that encloses an area $2\pi m\ell^2$, the ground state forms a QH droplet with area $2\pi N\ell^2$.

Many-body observables readily follow. Let us write $|\droplet\rangle$ for the QH ground state and $\hat c^{(\dagger)}(\bx)$ for fermionic annihilation (creation) operators. Then, for instance, the droplet's density
\beq
\rho(\bx)
\equiv
\langle\Omega|\hat c^{\dagger}(\bx)\hat c(\bx)|\Omega\rangle
=
\sum_{m=0}^{N-1}|\psi_m(\bx)|^2
\eeq
is found to be uniform in the bulk ($2\pi\ell^2 \ll \action \ll 2\pi N\ell^2$), where it takes the value $1/2\pi\ell^2$ consistent with the St\v{r}eda formula for a QH droplet with filling $\nu=1$. By contrast, near the edge ($\action \sim 2\pi N\ell^2$), the density falls off as an error function with width $\ell$. Another key observable is the electronic correlation function
\beq
\label{cxy}
C(\bx,\by)
\equiv
\langle\Omega|\hat c^{\dagger}(\bx)\hat c(\by)|\Omega\rangle
=
\sum_{m=0}^{N-1}\psi^*_m(\bx)\psi_m(\by),
\eeq
which is similarly found to be short-ranged in the bulk [recall Eq.~\eqref{t15x}] but long-ranged along the edge. 
Specifically, plugging the wave functions \eqref{psi_m} into Eq.~\eqref{cxy} and summing over $k\equiv N-m$ yields a geometric series due to the leading phase $\ee^{\ii m\theta}$. This results in the following asymptotic expression, verified in \cite{OLMSE:2024} for edge-deformed droplets:
\beq
C(\bx,\by)\big|_{\textrm{edge}}
\sim
\frac
  {\ee^{\ii\Theta(\bx,\by)}}
  {2\pi^{3/2} \ell}
\frac
  {\omega_{\textrm{F}}
  }
  {\sqrt{v(\theta_{\bx}) v(\theta_{\by})}}
\frac
  {\ii\exp \left( -\frac{d_{\bx}^2}{2\ell^2}-\frac{d_{\by}^2}{2\ell^2} \right)}
  {2\sin\bigl( [\theta_{\bx} - \theta_{\by}]/2 \bigr)}.
\label{eq:C_edge}
\eeq
Here, $\Theta(\bx_1,\bx_2)$ is a complicated overall phase, $\omega_{\textrm{F}}\equiv\dot\theta(\action_N)$ is the angular Fermi velocity, $v(\theta)\equiv v_N(\theta)$ is the norm \eqref{v_theta} of the edge velocity, and we assume $\theta_{\bx} \neq \theta_{\by}$. We have thus recovered Eq.~\eqref{cedge} in the main text, with the normalization $\mathcal{N} \equiv \omega_{\textrm{F}}/2\pi^{3/2}\ell$. Finally, the connected density-density correlator for free fermions is
\beq
\langle\Omega|\hat\rho(\bx)\hat\rho(\by)|\Omega\rangle_c
\equiv
\langle\Omega|\hat c^{\dagger}(\bx)\hat c(\bx)\hat c^{\dagger}(\by)\hat c(\by)|\Omega\rangle-\rho(\bx)\rho(\by)
=
\rho(\bx)\delta(\bx-\by)-|C(\bx,\by)|^2,
\eeq
incidentally implying Eq.~\eqref{s15x} in the main text. It follows that density-density correlations along the QH edge are given by the square of Eq.~\eqref{eq:C_edge} provided $\theta_{\bx}\neq\theta_{\by}$:
\beq
\label{e40b}
\langle\Omega|\hat\rho(\bx)\hat\rho(\by)|\Omega\rangle_c\big|_{\text{edge}}
\sim
-\frac{1}{4\pi^3\ell^2}
\frac
  {\omega_{\textrm{F}}^2}
  {v(\theta_{\bx}) v(\theta_{\by})}
\frac
  {\exp\left( -\frac{d_{\bx}^2}{\ell^2}-\frac{d_{\by}^2}{\ell^2} \right)}
  {4\sin^2\bigl( [\theta_{\bx} - \theta_{\by}]/2 \bigr)}.
\eeq
Neglecting all prefactors on the right-hand side, constant or angle-dependent, leaves the leading behavior \eqref{s15h} announced in the main text. As stressed there, the key property of this result is its dependence on the canonical angle coordinate $\theta$, as opposed to other conceivable coordinates along the QH edge. Note that the omission of prefactors can be made more precise, as the quantity \eqref{s15h} is the actual density-density correlator of the edge conformal field theory (CFT). Indeed, we showed in \cite{OLMSE:2024} that a double (Gaussian) integral of Eq.~\eqref{eq:C_edge} along the action variables of $\bx,\by$ gives rise to the universal $1/\sin\bigl( [\theta_{\bx} - \theta_{\by}]/2 \bigr)$ correlator of chiral fermions. Its square is the density-density correlator in the edge CFT, whose expression is just Eq.~\eqref{s15h} without any influence of the local velocity \eqref{v_theta}.

\section*{Part~B: Coulomb gases}

Let us now review the physics of classical 2D CGs. This material is well understood in the literature, but we cover it here for completeness. The key formula is Eq.~\eqref{eq:confedge}, which is repeatedly used in the main text. For more, see \eg \cite{Jancovici:1995, Jancovici:1982ii, AlastueyJancovici:1984, ChoquardEtAl:1987, Dyson:1962}.

\subsection*{A classical conductor at finite temperature}

Consider a gas of $N$ classical point charges in the plane, with positions $\bx_j\in\mathbb{R}^2$, $j=1,\ldots,N$. Given a configuration $\bx_1,\ldots,\bx_N$, we assume as in Eq.~\eqref{s1o} that its energy is 
\beq
\label{eq:energy}
E(\bx_1,\ldots,\bx_N)
\equiv
- \sum_{1\leq i<j\leq N}\log |\bx_i-\bx_j| + \frac{N}{2}\sum_{j=1}^N W(\bx_j),
\eeq
where the first term is a 2D electrostatic interaction and the second reflects the presence of some (real) external potential $W$. The latter is assumed to grow sufficiently fast at infinity $|\bx|\to\infty$ (anything faster than $\log |\bx|$ is sufficient). Note also the factor $N$ in front of the last term, ensuring that interactions and potential are of the same order if the $|\bx_j|$ are all of order one, and facilitating comparison with some of the literature.

As defined in Eq.~\eqref{eq:energy}, the energy of the system is dimensionless for simplicity, which is to say that its temperature is measured in units of some (unimportant) overall energy scale. Indeed, we are interested in placing this system in the canonical ensemble at temperature $\beta^{-1}$. Then the joint probability density of finding the particles at positions $\bx_1, \ldots, \bx_N$ is 
\beq
\label{eq:jpdf}
P(\bx_1,\ldots,\bx_N)
=
\frac{\ee^{-\beta E(\bx_1,\ldots,\bx_N)}}{Z}
\qquad\text{with}\qquad
Z
\equiv
\int \dd\bx_1 \ldots \dd\bx_N\, \ee^{-\beta E(\bx_1,\ldots,\bx_N)},
\eeq
where $\dd \bx_j \equiv \dd x_j \dd y_j$ and $Z$ is the usual canonical partition function. This readily leads to the plasma analogy that relates CGs to QH droplets. Indeed, the probability density \eqref{eq:jpdf} coincides with the probability density of an isotropic Laughlin wave function of $N$ electrons at filling fraction $\nu = 2/\beta$ \cite{Laughlin:1983anomalous}, provided the CG potential is $W(\bx)=|\bx|^2$ in terms of dimensionless Cartesian coordinates $\bx\equiv{\bf X}/\sqrt{\beta N\ell^2}$, where $\ell$ is the magnetic length \eqref{eldef} and ${\bf X}$ are dimensionful Cartesian coordinates. The value $\beta=2$ corresponds to the free-fermion point, where the QH droplet is a Slater determinant of wave functions in the LLL.

The study of CGs essentially consists in computing expectation values of various local observables in the thermodynamic limit $N\to\infty$. Given any observable $\cO(\bx_1,\ldots,\bx_N)$, its mean and variance are respectively
\beq
\langle\cO\rangle
\equiv
\int\dd\bx_1 \ldots \dd\bx_N
\cO(\bx_1,\ldots,\bx_N) P(\bx_1,\ldots,\bx_N),
\qquad
\Var[\cO]
\equiv
\langle\cO^2\rangle-\langle\cO\rangle^2.
\eeq
For instance, the particle density introduced above Eq.~\eqref{e7} is the random variable $\frac{1}{N}\sum_{j=1}^N \delta(\bx-\bx_j)$, and its mean is
\beq
\label{e45}
\rho(\bx)
\equiv
\frac{1}{N}\!\left\langle\sum_{j=1}^N\delta(\bx-\bx_j)\right\rangle\!.
\eeq
This is normalized so that $\int \dd\bx\,\rho(\bx) = 1$, since the total number $N$ of particles does not fluctuate. For an isotropic harmonic potential $W(\bx)=|\bx|^2$, the density \eqref{e45} in the thermodynamic limit is uniform in a disk and zero outside. In what follows, we explain this fact and various others, \eg involving the density-density correlator $\bigl\langle\sum_{i,j}\delta(\bx-\bx_i)\delta(\bx-\bx_j)\bigr\rangle$ or the variance of the one-body observable $\sum_{j=1}^N \LSFunc(\bx_j)$ for some (smooth) function $\LSFunc$.

\subsection*{Energy minimization and resulting droplet}

Let us ask how the mean density \eqref{e45} behaves in the limit $N\to\infty$. In statistical mechanics, there is usually a nontrivial competition between energy and entropy, but this is not so here (at least to leading order) due to long-range Coulomb interactions: the energy \eqref{eq:energy} is of order $N^2$, while entropic fluctuations turn out to be of order $N$ \cite{Dyson:1962}. The density profile can thus be found from mere energy minimization, without thermal effects. For large $N$, one can approximate sums by integrals to rewrite the energy \eqref{eq:energy} as $E \approx (N^2/2) \mathcal{E}[\mu]$, where what one seeks to minimize is the energy functional 
\beq
\label{e47}
\mathcal{E}[\mu]
=
- \int \dd\bx \dd\by\, \mu(\bx) \mu(\by)\log|\bx-\by|
  + \int \dd\bx\, \mu(\bx) W(\bx).
\eeq
The minimizer $\mu=\rho$ is to be found among all possible densities $\mu$, \ie nonnegative functions that integrate to one, $\int \dd\bx\, \mu(\bx) = 1$. Under mild technical conditions on the potential $W$, one can show that the minimizer exists, is unique, and has compact support; see \cite{SaffTotik:1997} for a mathematical discussion. The region where $\rho$ is supported is identified as the droplet $\droplet$; then $\rho(\bx)=0$ for all $\bx\notin\droplet$.

From a field-theoretic perspective, it is straightforward to find the minimizer $\rho$ of \eqref{e47}. One can indeed deal with the normalization constraint by introducing a Lagrange multiplier $\lambda$ and defining the new functional
\beq
\mathcal{E}'[\mu,\lambda]
\equiv
- \int \dd\bx \dd\by\, \mu(\bx)\mu(\by) \log |\bx-\by|
  + \int \dd\bx\, \mu(\bx)W(\bx)
  - \lambda \left[ \int \dd\bx\, \mu(\bx) - 1 \right].
\eeq
The corresponding Euler-Lagrange equations are $\partial\mathcal{E}'/\partial \lambda=0$ and $\delta \mathcal{E}' / \delta \mu(\bx) = 0$. The first reproduces the normalization constraint, while the second implies that $\rho$ satisfies the screening condition \eqref{e7}:
\beq
\label{eq:funda}
W(\bx) - \int_{\droplet} \dd\by\, \rho(\by)\log |\bx-\by|^2 = \lambda
\quad \forall\,\bx \in \droplet,
\eeq
for some fixed $\lambda$. By inspection, one can infer why the droplet must be finite: the integral in Eq.~\eqref{eq:funda} grows logarithmically as $|\bx|\to\infty$ while the potential $W(\bx)$ grows faster than that, so the equality cannot hold if $\droplet$ has noncompact support. Now, taking the Laplacian $\nabla^2 = \partial_x^2 + \partial_y^2$ of both sides of Eq.~\eqref{eq:funda} gives
\beq
\label{eq:laplacian}
\rho(\bx) = \frac{\nabla^2 W(\bx)}{4\pi}
\quad \forall\,\bx \in \droplet,
\eeq
which was written in the main text below Eq.~\eqref{e7}. Note that we implicitly assume the Laplacian of $W$ to be positive, which is not a restriction for the plasma analogy where the bulk density is uniform anyway. For the example $W(\bx)=|\bx|^2$, it follows that $\rho(\bx)=1/\pi$ inside the droplet $\droplet$, which is a unit disk.

Plugging Eq.~\eqref{eq:laplacian} back into the screening condition \eqref{eq:funda} implicitly determines the droplet $\droplet$. In practice, finding $\droplet$ in this way is a difficult inverse problem; the most one can say at first glance is that anisotropic potentials yield anisotropic droplets. However, one can reverse the logic by first choosing a region $\droplet$ and a density $\rho$ therein, then asking what potential $W$ corresponds to that choice. The answer is trivially provided by the screening condition \eqref{eq:funda}, which yields
\beq
\label{eq:potchoice}
W(\bx)
=
\int_{\droplet} \dd\by\, \rho(\by)\log |\bx-\by|^2
\qquad\text{for }\bx\in\droplet.
\eeq
Outside of $\droplet$, the potential is nearly unconstrained: all that is needed is that $W(\bx)$ be greater than $\max_{\bx\in \droplet} |W(\bx)|$, with sufficiently fast growth at infinity. It is even possible to take $W(\bx) = +\infty$ outside to ensure this, since the potential need not be continuous.

Eq.~\eqref{eq:potchoice} is useful to numerically simulate CGs of any given shape. For comparison with QH droplets, we restrict attention to CGs with uniform bulk density \eqref{eq:laplacian}, \ie CGs whose potential has constant Laplacian; see \eg \cite{ZabrodinWiegmann:2006} for a discussion. Without loss of generality, we choose the droplet's area to equal $\pi$, hence the normalized bulk density $\rho=1/\pi$. It is then convenient to use complex coordinates $z\equiv x^1+\ii x^2$, $w\equiv y^1+\ii y^2$ and turn the surface integral \eqref{eq:potchoice} into a line integral thanks to Stokes' theorem:
\beq
\label{2d1d}
W(z)=|z|^2+\textrm{Re}\left[\oint_{\partial \Omega} \frac{\dd w}{i \pi}\bar{w}\log(1-z/w)\right].
\eeq
Eq.~\eqref{eq:potchoice} guarantees that this holds for $z\in\droplet$, but it also turns out to be a valid choice outside the droplet ($z\notin\droplet$). The key advantage of Eq.~\eqref{2d1d} is that it can be approximated numerically to high precision by a rectangle rule, so it provides a good approximation of the potential needed to simulate any given droplet shape. In the main text, Eq.~\eqref{2d1d} was used to produce the potential corresponding to flower droplets \eqref{eq:edge}. The potential was then plugged into the energy \eqref{eq:energy} to eventually yield the Monte Carlo data behind Fig.~\ref{fig5}, similar to that in \cite{CardosoEtAl:2021}.

\subsection*{Correlations in Coulomb gases}

The left-hand side of the screening condition \eqref{eq:funda} is the total electric potential in the bulk of a CG, due to the external potential $W(\bx)$ and to the potential created by the CG itself. Eq.~\eqref{eq:funda} thus says that the bulk of a CG supports a vanishing electric field, \ie that a CG is a perfect 2D conductor. It follows that bulk perturbations are screened by charges which rearrange themselves so that the bulk potential \eqref{eq:funda} remains constant, while perturbations outside of $\droplet$ have genuinely long-range effects. This is illustrated in Fig.~\ref{fig:screening}, and it leads to charge accumulation at the CG boundary, as for any conductor. In turn, this implies sharply different behaviors for bulk and edge correlations of point charges. Such correlations are essential for our argument on the mismatch between QH droplets and CGs, so we review them here.

As explained in the main text, the key object for correlations is the field \eqref{eq:potfield}, \ie the potential
\beq
\label{s145b}
\Phi(\bx)
\equiv
-\sum_{j=1}^N\log|\bx-\bx_j|
\eeq
created at $\bx$ by all the point charges of the CG. Its mean value in the thermodynamic limit is
\beq
\label{e50}
\lim_{N\to\infty}\frac{1}{N}\langle\Phi(\bx)\rangle
=
-\int_{\Omega}\dd\by\, \rho(\by)\log|\bx-\by|,
\eeq
so it is indeed the potential created by the CG in the screening condition \eqref{eq:funda}. To see how correlations come about, consider the effect of adding a small point charge $q$ at $\bx$ to the gas with energy \eqref{eq:energy}. Then the perturbed energy is $E+q\Phi(\bx)$ and the mean of the potential \eqref{s145b} in the perturbed CG is
\beq
\label{t145b}
\langle\Phi(\by)\rangle^{\bx}
\equiv
\frac{\int\dd\bx_1\ldots\dd\bx_N\, \Phi(\by)\ee^{-\beta E-\beta q\Phi(\bx)}}{\int\dd\bx_1\ldots\dd\bx_N\, \ee^{-\beta E-\beta q\Phi(\bx)}}
=
\langle\Phi(\by)\rangle
-q\beta\underbrace{\Bigl( \langle\Phi(\bx)\Phi(\by)\rangle - \langle\Phi(\bx)\rangle \langle\Phi(\by)\rangle \Bigr)}_{\displaystyle\equiv\langle\Phi(\bx)\Phi(\by)\rangle_c} +\, O(q^2).
\eeq
Here, we expanded the mean up to first order in $q$, \ie at the order of linear response, and we introduced the connected two-point function of $\Phi$. The result \eqref{t145b} was announced in the main text in Eq.~\eqref{e88}. It says that correlations of the potential \eqref{s145b} determine the effects of small perturbations of the CG.

One can now combine Eq.~\eqref{t145b} with screening to determine the universal behaviors of the correlator $\langle\Phi(\bx)\Phi(\by)\rangle_c$ in the thermodynamic limit \cite{Jancovici:1995}. The answer depends on whether the points $\bx,\by$ are inside or outside the conductor. In total, there are three different cases:
\begin{itemize}
\setlength\itemsep{0em}
\item
Assume that both $\bx$ and $\by$ are in the bulk, though not too close to each other. By screening \eqref{eq:funda}, the total potential $\langle\Phi(\by)\rangle^{\bx}-\langle\Phi(\by)\rangle-q\log|\bx-\by|$ is constant, whereupon Eq.~\eqref{t145b} implies
\beq
\label{s14t}
\langle\Phi(\bx)\Phi(\by)\rangle_c
=
-\frac{1}{\beta}\log|\bx-\by|+\text{const}
\qquad
\text{for }\bx,\by\in\droplet.
\eeq
In this sense, the potential \eqref{s145b} behaves as a massless free boson in the bulk of a CG. We stress, however, that the relevant conformal structure will be \emph{outside} the droplet, and that it ultimately has nothing to do with the conformal behavior of $\Phi(\bx)$ as a free boson.

\item
Now let $\bx$ be outside the droplet while $\by$ lies in the bulk. Then the total potential $\langle\Phi(\by)\rangle^{\bx}-\langle\Phi(\by)\rangle-q\log|\bx-\by|$ is that created by a perfect conductor with $N+1$ particles, and it is therefore harmonic when seen as a function of $\bx$. Hence,
\beq
\label{s14tbis}
\langle\Phi(\bx)\Phi(\by)\rangle_c
=
-\frac{1}{\beta}\log|\bx-\by|+H(\bx)
\qquad
\text{for }\bx\notin\droplet,\,\by\in\droplet,
\eeq
where $H$ is harmonic. The latter only depends on $\bx$, which is ultimately why it will play no role for edge correlations below. The case where $\bx$ is in the bulk while $\by$ is outside is analogous, as follows from the symmetry of the two-point correlator.

\item
Finally, let both $\bx$ and $\by$ lie outside the droplet. Then the total potential $\langle\Phi(\by)\rangle^{\bx}-\langle\Phi(\by)\rangle-q\log|\bx-\by|$ is the Green's function for the Laplacian $-(1/q)\nabla^2$ outside $\droplet$, with Dirichlet boundary conditions on the edge $\partial \droplet$. Thus,
\beq
\label{s145t}
\langle\Phi(\bx)\Phi(\by)\rangle_c
= - \frac{1}{\beta}\log|\bx-\by|
  - \frac{1}{\beta}\greensFunc(\bx,\by)
  + \text{const}
\qquad
\text{for }\bx,\by\notin\droplet,
\eeq
where $\greensFunc(\bx,\by)$ satisfies $-\nabla_{\bx}^2 \greensFunc(\bx,\by)=2\pi \delta(\bx-\by)$. Such Green's function are most conveniently studied using complex analysis, so in the following we identify vectors $\bx,\by \in \mathbb{R}^2$ with complex numbers $z,w\in \mathbb{C}$. If the droplet is the unit disk, the Green's function can be computed through elementary means and reads
\beq
\greensFunc_0(z,w)
=
\log \left|\frac{1-z \bar{w}}{z-w}\right|.
\eeq
For more complicated simply connected domains, the Green's function can be expressed using conformal maps; see \eg \cite{SaffTotik:1997}. More precisely, it is given by
\beq
\label{e56}
\greensFunc(z,w)
=
\greensFunc_0\big(G(z),G(w)\big)
=
\log \left|\frac{1-G(z)\overline{G(w)}}{G(z)-G(w)}\right|,
\eeq
where $G$ is the (unique) conformal map \eqref{ss15n}, from the exterior of the droplet $\droplet$ to the exterior of the unit disk, that behaves at infinity as $G(z)\sim z/a$ for some $a>0$. We stress that ``conformal'' here means that the map \eqref{ss15n} is a biholomorphism (a holomorphic bijection).
\end{itemize}

Since the map \eqref{ss15n} sends $\partial\droplet$ on a unit circle, one can label any point $\bx\in\partial\droplet$ by the angle $\alpha$ such that $G(\bx)=\ee^{\ii\alpha}$. This is the conformal angle coordinate that governs the density-density correlations \eqref{s15n} on the edge of a CG. To derive these correlations, the key intuition stems again from electrostatics: as we saw, correlators of the field \eqref{s145b} are closely related to the electric potential generated at $\by$ (or $w$) by a point charge set at $\bx$ (or $z$) in the presence of a conductor, the droplet $\droplet$. The resulting electric field is discontinuous on the boundary $\partial\droplet$ (see Fig.~\ref{fig:screening}), resulting in a nonzero edge charge density $\sigma(\bx)$ for $\bx\in\partial\droplet$. The main argument in \cite{Jancovici:1995} is that this phenomenon is responsible for the power-law decay of edge density correlations. Indeed, the edge charge density is
\beq
\sigma(\bx)
\equiv
\frac{1}{2\pi}\mathbf{n}_\bx\cdot\bigl[\mathbf{E}_+(\bx)-\mathbf{E}_-(\bx)\bigr],
\label{s155}
\eeq
where $\mathbf{n}_\bx$ is the outward normal unit vector at $\bx\in\partial\droplet$ and $\mathbf{E}_{\pm}(\bx)\equiv\mathbf{E}(\bx\pm\epsilon\mathbf{n}_{\bx})$ are electric fields right outside and right inside the droplet. Since the electric field is $\mathbf{E}=-\nabla\Phi$, correlations of the charge density \eqref{s155} can be related to those of $\Phi$ in Eqs.~\eqref{s14t}--\eqref{s145t}. The only nonzero contribution to the correlator $\langle\sigma(\bx)\sigma(\by)\rangle_c$ turns out to stem from the Green's function $K(\bx,\by)$ in Eq.~\eqref{s145t}, and one finds
\beq
\braket{\sigma(\bx)\sigma(\by)}_c
=
-\frac{1}{4\pi^2\beta}
\nabla_{\mathbf{n}_\bx}\nabla_{\mathbf{n}_\by}\greensFunc(\bx,\by)
\qquad
\text{for }\bx,\by\in\partial\droplet.
\eeq
Here, we assume that $\bx,\by$ do not coincide, and the derivatives are taken in a direction normal to the boundary. The result can be recast in complex coordinates using the map \eqref{ss15n} and the Green's function \eqref{e56}, which yields
\beq
\label{eq:confedge}
\braket{\sigma(z)\sigma(w)}_c
=
-\frac{1}{2\pi^2\beta}\frac{|G'(z)G'(w)|}{|G(z)-G(w)|^2}
\qquad
\text{for }z,w\in\partial\droplet.
\eeq
This is the key formula for edge correlations in a CG: using the angle such that $G(z)=\ee^{\ii\alpha}$ on $\partial\droplet$, the denominator of Eq.~\eqref{eq:confedge} is that announced in Eq.~\eqref{s15n} of the main text. Note the striking similarity with the QH correlator \eqref{e40b} integrated in a direction normal to the boundary, with factors $1/v(\theta)$ playing a role analogous to the derivatives $G'(z)$ in Eq.~\eqref{eq:confedge}. In fact, at the free fermions point $\beta=2$, Eq.~\eqref{eq:confedge} can be recovered from an asymptotic expansion of microscopic correlations similar to those of QH droplets in Eq.~\eqref{e40b}; see Theorem~1.3 in \cite{AmeurCronvall:2023}.

A comment is in order regarding the conformal map \eqref{ss15n} that plays a central role in the correlator \eqref{eq:confedge}. It turns out that the inverse map $G^{-1}:\mathbb{R}^2\backslash\mathbb{D}\to\mathbb{R}^2\backslash\droplet$ is often slightly easier to understand than the actual map itself, and that it can always be written as
\beq
G^{-1}(\zeta)=a\zeta+\sum_{k\geq 0} a_k \zeta^{-k}
\eeq
for some complex coefficients $a$ and $a_k$ ($k \in \mathbb{N}$). The latter are related via $|a|^2=1+\sum_{k\geq 1} k|a_k|^2$ when $\mathbb{D}$ is a unit disk and the droplet $\droplet$ has area $\pi$ (without loss of generality). In practice, analytical expressions for conformal maps are rare, so one often has to rely on numerical evaluations, based \eg on Fekete points \cite{SaffTotik:1997}. A simple exception is the case of an elliptic droplet, for which the inverse conformal map is
\beq
\label{e60}
G^{-1}(\zeta)=\zeta\cosh(\lambda)+\frac{1}{\zeta}\sinh(\lambda),
\eeq 
where the half-width is $e^{\lambda}$ and the half-height is $e^{-\lambda}$. Another is the square droplet (or more generally any polygonal droplet \cite{DriscollTrefethen:2002}), for which the inverse of \eqref{ss15n} involves the Gaussian hypergeometric function:
\beq
\label{eq:confsquare}
G^{-1}(\zeta)
= \frac{\Gamma(1/4)\,\zeta}{2\sqrt{2}\Gamma(3/4)} \,_{2}F_1(-1/2,-1/4;3/4;-1/\zeta^4).
\eeq
Eqs.~\eqref{e60} and \eqref{eq:confsquare} were used in Fig.~\ref{fig3} of the main text, respectively for the ellipse and the square.

\subsection*{Fluctuations of one-body observables}

We conclude our review of CGs with the fluctuations of one-body observables $\sum_j U(\bx_j)$, mentioned in Sec.~\ref{se5} of the main text and also known as linear statistics in the CG literature. In particular, we recall how Eq.~\eqref{t15w} follows from screening and the edge correlator \eqref{eq:confedge}.

Our starting point is to express the variance through density-density correlations, similarly to Eq.~\eqref{s15x}:
\beq
\label{e62}
\textrm{Var}[U]_{\text{CG}}
=
-\frac{1}{2} \int \dd\bx\,\dd\by \left[U(\bx)-U(\by)\right]^2\braket{\rho(\bx)\rho(\by)}_c,
\eeq
where $\dd\bx\equiv\dd x^1\dd x^2$ and similarly for $\by$. The strategy will be to split the integral on the right-hand side in two pieces---one due to short-range bulk correlations, the other to long-range edge correlations. At leading order in the thermodynamic limit, the two contributions decouple. To compute the bulk term, it is easiest to first consider a one-body observable $\tilde{U}$ which vanishes on $\partial \droplet$. Bulk correlations are translation-invariant, so going to center-of-mass coordinates $\br\equiv(\bx+\by)/2$ and $\bs\equiv\bx-\by$ yields
\beq
\textrm{Var}[\tilde{U}]_{\text{CG}}
=
- \frac{1}{2}\int_{\droplet} \dd\br \int \dd\bs\, \Bigl[ \tilde{U}(\br+\bs/2) - \tilde{U}(\br-\bs/2) \Bigr]^2 \! \braket{\rho(\bs)\rho(0)}_c
\sim
- \frac{1}{2}\int_{\droplet} \dd\br \int \dd\bs\, \Bigl[ \bs \cdot \nabla\tilde{U}(\br) \Bigr]^2 \! \braket{\rho(\bs)\rho(0)}_c,
\eeq
where we used the fast decay of correlations to perform a Taylor expansion of $\tilde{U}$. Further using the rotational invariance of bulk correlations yields the desired result,
\beq
\textrm{Var}[\tilde{U}]_{\text{CG}}
\sim
-\frac{1}{4}\int_{\droplet} \dd\br\, (\nabla \tilde{U})^2 \int \dd\bs \,\bs^2 \braket{\rho(\bs)\rho(0)}_c
=
\int_{\droplet}\frac{\dd\br}{2\pi\beta}|\nabla \tilde{U}|^2,
\label{e65b}
\eeq
where the second equality follows from the Stillinger-Lovett sum rule \cite{StillingerLovett:1968}
\beq
\int \dd\bs \,\bs^2 \braket{\rho(\bs)\rho(0)}_c
=
-\frac{2}{\pi\beta}.
\eeq 
We emphasize that the latter is exact for all $\beta$, even though the two-point function itself is not known away from the free-fermion point $\beta=2$. Formally, the sum rule follows from $\nabla^2 \Phi(\bx)=-2\pi\rho(\bx)$, writing $\braket{\rho(\bx)\rho(\by)}_c=\nabla^2_{\bx}\nabla^2_{\by}\braket{\Phi(\bx)\Phi(\by)}_c$, using the bulk result \eqref{s14t} and performing integration by parts. It can thus be seen as a sum rule for the density of any perfect 2D conductor.

For general $U$ as opposed to $\tilde U$, edge effects come into play. A reasoning similar to the one that led to Eq.~\eqref{e65b} then yields
\beq
\label{e67b}
\textrm{Var}[U]_{\text{CG}}
\sim
\int_{\droplet} \frac{\dd\bx}{2\pi\beta} |\nabla U(\bx)|^2
-\frac{1}{2} \oint_{\partial\droplet} \frac{\dd z}{2\pi iz}\oint_{\partial\droplet}\frac{\dd w}{2\pi iw}[U(z)-U(w)]^2 \braket{\sigma(z)\sigma(w)}_c.
\eeq
Here, the edge correlator $\braket{\sigma(z)\sigma(w)}_c$ appears due to the double integral, in Eq.~\eqref{e62}, along a direction perpendicular to the boundary $\partial\droplet$. Eq.~\eqref{t15w} of the main text then follows from Eq.~\eqref{e67b} upon using the edge correlator \eqref{eq:confedge}. While physically reasonable, the decoupling between bulk and edge contributions in Eq.~\eqref{e67b} can also be checked by explicit computations at the free-fermion point $\beta=2$. It is ultimately justified by the fact that Eq.~\eqref{t15w} can be proved with more advanced techniques; see \cite{LebleSerfaty:2018}.

Finally, note that Eqs.~\eqref{t15w} and \eqref{e67b} can be recast as
\beq
\label{eq:varianceconfbis}
\textrm{Var}[U]_{\text{CG}}
\sim
\int_{\droplet} \frac{d^2z}{2\pi\beta} |\nabla U|^2+\frac{2}{\beta} \sum_{k=1}^{\infty} k U_k U_{-k},
\eeq
where $U_k=\int_0^{2\pi}\frac{\dd\alpha}{2\pi}U(G^{-1}(\ee^{\ii \alpha}))\ee^{-\ii k \alpha}$ is the $k^{\text{th}}$ Fourier mode of the periodic function $U(G^{-1}(\ee^{\ii \alpha}))$. This alternative form is sometimes convenient for concrete calculations. For instance, the variance of $U(z)=\textrm{Re}(z)$ for a square droplet follows from the conformal map \eqref{eq:confsquare} and reads
\beq
\textrm{Var}[U]_{\text{CG}}
=
\frac{1}{2\beta}+\frac{\Gamma(1/4)^2}{192\beta\, \Gamma(3/4)^2}\left[12+\, \!_{4}F_3\left(\frac{1}{2},\frac{1}{2},\frac{3}{4},\frac{3}{4};\frac{7}{4},\frac{7}{4},2;1\right)\right]\
\eeq
in terms of a generalized hypergeometric function. This is the exact formula behind the second value in Eq.~\eqref{VarU_QHCG} of the main text. It should be contrasted (at $\beta=2$) with the QH variance \eqref{cedge} for an edge-deformed square droplet: the QH variance can then be written as an infinite series
\beq
\textrm{Var}[\hat{U}]_{\textrm{QH}}
=
\frac{1}{4}+\sum_{m=0}^{\infty} (2m+1)\left|\int_{-\pi/4}^{\pi/4}8\sqrt{\pi} e^{-i (-1)^m\varphi}\frac{(\pi+4i \varphi)^{m-1}}{(\pi-4i \varphi)^{m+2}}d\varphi\right|^2,
\eeq
whose numerical value is the first one given in Eq.~\eqref{VarU_QHCG}.

\section*{Part~C: Symplectic vs conformal maps}

Since edge correlations of QH droplets and CGs are governed by the angles \eqref{s1da}--\eqref{s1db}, a natural question is whether these sometimes coincide. An equivalent phrasing is to ask if the restriction of the area-preserving map \eqref{t15s} to the edge $\partial\droplet$ can be extended to a biholomorphism (a holomorphic bijection) outside the droplet $\droplet$.

Here, we address this puzzle within the class of edge-deformed potentials \eqref{s1e}, with special focus on flower droplets. The corresponding action-angle coordinates are given by Eq.~\eqref{ssc} with $R=0$ and a function $\theta(\varphi)$ given by Eq.~\eqref{s1z}. As we shall see, an extension from the canonical angle to a biholomorphism is possible for elliptic droplets ($k=2$), but impossible for droplets with $k\neq2$ petals. Despite this, we will find that the local (not globally bijective) conformal extension does replicate the correlator of edge-deformed QH droplets.

\subsection*{Statement of the problem}

Let $F$ be the area-preserving map \eqref{t15s} sending a droplet $\droplet$ on a disk $\mathbb{D}$. The restriction of $F$ to the edge $\partial\droplet$ sends it on a circle $S^1=\partial\mathbb{D}$, which defines the map $F_{\partial}\equiv\theta$ in Eq.~\eqref{s1da}. Similarly, let $G$ be the biholomorphism \eqref{ss15n} that conformally maps the outside of $\Omega$ on that of $\mathbb{D}$, and let $G_{\partial}\equiv\alpha$ in Eq.~\eqref{s1db} be its edge restriction. The question is:
\begin{enumerate}[label = {\bf Q}:, ref = {\bf Q}]
\item
\label{ss16b}
Given an area-preserving map $F$, is there a biholomorphism $G$ such that $F_{\partial} = G_{\partial}$?
\end{enumerate}
We shall address this problem in the specific case where $F$ is an edge deformation, as defined around Eq.~\eqref{s1e} in the main text. Any such map takes the following form in polar coordinates:
\beq
F(r,\varphi)
\equiv
\bigg(\frac{r}{\sqrt{f'(\varphi)}},f(\varphi)\bigg),
\label{s16b}
\eeq
where $f(\varphi)$ is a circle diffeomorphism that satisfies the conditions \eqref{tc}. [To avoid confusion with the map \eqref{s1da}, we denote the circle diffeomorphism by $f(\varphi)$ rather than $\theta(\varphi)$.] Since the edge $\partial\droplet$ is the set of points that satisfy Eq.~\eqref{eq:edge} in the main text, with some fixed area $\action$, the edge restriction of the map in Eq.~\eqref{s16b} is
\beq
F_{\partial}:
\partial\droplet\to S^1:
\bigg(r=\sqrt{\frac{\action}{\pi}f'(\varphi)},\varphi\bigg)\mapsto\bigg(\sqrt{\frac{\action}{\pi}},f(\varphi)\bigg).
\label{t16b}
\eeq
The question is whether this can be extended to a biholomorphism \eqref{ss15n}. In that context, the area $\action$ does not matter (it can be rescaled by a global dilation), and it is more convenient to seek to extend the \emph{inverse} of Eq.~\eqref{t16b},
\beq
F_{\partial}^{-1}:
S^1\to\partial\droplet:
\ee^{\ii\theta}
\mapsto
\frac{\ee^{\ii f^{-1}(\theta)}}{\sqrt{(f^{-1})'(\theta)}},
\label{s15t}
\eeq
where $f^{-1}$ denotes the inverse of the function $f$ in Eq.~\eqref{t16b} and we expressed everything in terms of $\theta = f(\varphi)$. The right-hand side of \eqref{s15t} can be written as $\ee^{\ii\theta}P(\theta)$, with $P(\theta)$ some complex, smooth $2\pi$-periodic function that admits a Fourier expansion. It follows that the local conformal extension of the map \eqref{s15t} is unique: it is obtained by viewing $F_{\partial}^{-1}$ as the restriction to $S^1$ of
\beq
H^{-1}(\zeta)
\equiv
\frac{\ee^{\ii f^{-1}(-\ii\log \zeta)}}{\sqrt{(f^{-1})'(-\ii\log \zeta)}},
\label{t15t}
\eeq
where $\zeta\equiv x+\ii y$ is the usual complex coordinate in the plane. Note that the argument of the function $f^{-1}$ is now complexified and involves a logarithm, suggesting that $H^{-1}$ generally has branch cuts or singularities. [Despite this, the right-hand side of Eq.~\eqref{t15t} can be understood as a formal Laurent series in $\zeta$ in a neighborhood of the circle $|\zeta| = 1$.] Indeed, we will see that precisely this issue is responsible for the absence of a biholomorphic extension of flower deformations with $k\neq2$ petals. This is also why we denote the local conformal map \eqref{t15t} by $H$ rather than $\conformal$ as in \eqref{ss15n}: in general, the maps $H$ and $\conformal$ differ. Only $\conformal$---not $H$---determines correlations such as \eqref{eq:confedge} in a CG.

We stress that the converse of Question~\ref{ss16b} is trivial: given any biholomorphism $\conformal$ in \eqref{ss15n} such that $\text{area}(\droplet) = \text{area}(\mathbb{D})$, it is always possible to find a smooth area-preserving map $\symplectic$ in \eqref{t15s} such that $\symplectic_{\partial}=\conformal_{\partial}$. Indeed, given any coordinate $\alpha$ along $\partial\droplet$, there certainly exists a smooth confining potential $V(\bx)$ that admits $\partial\droplet$ as a level curve, with a gradient $\nabla V$ whose norm \eqref{v_theta} along $\partial\droplet$ gives rise to $\alpha$ as the canonical coordinate \eqref{sa}. This is the sense in which any CG, of any shape, admits a QH realization provided the QH potential is suitably fine-tuned. But the opposite is not true: most QH droplets do \emph{not} admit a plasma analogue.

\subsection*{From symplectic to conformal flower deformations}

\begin{figure}[htbp]
\centering \includegraphics[width=0.85\textwidth]{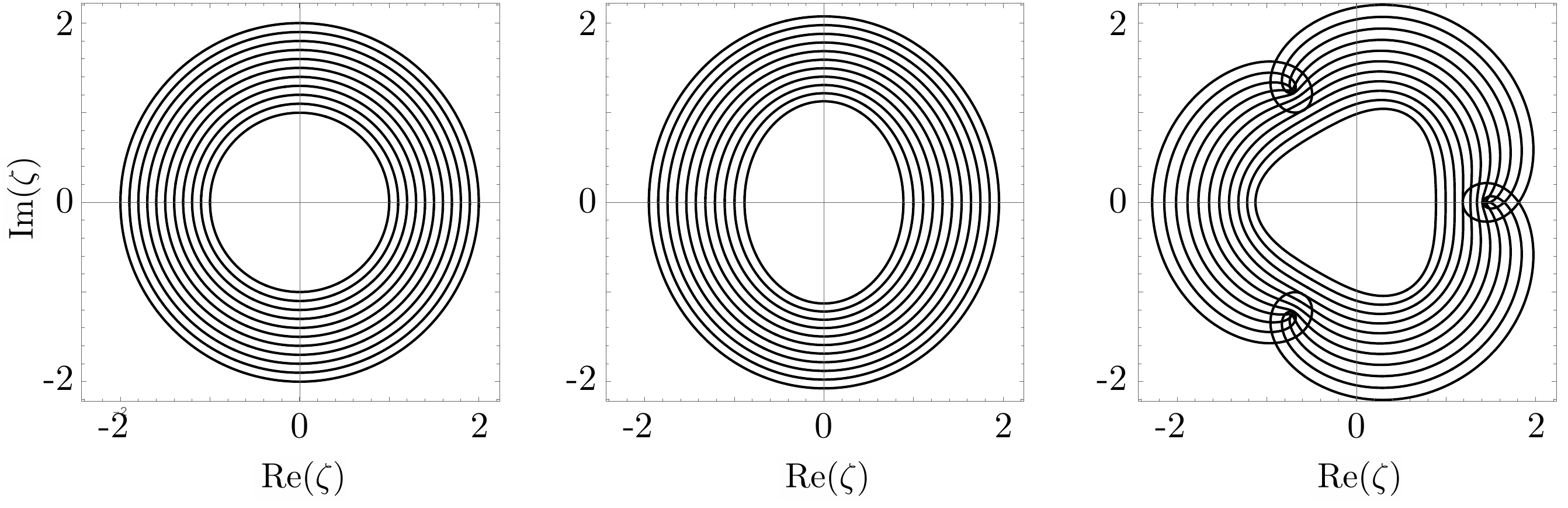}
\caption{Left: Exterior of the unit disk; only the annulus with outer radius $\leq 2$ is shown for convenience. Middle and Right: Images of the annulus under the map \eqref{e67} with $\lambda=0.12$, $k=2$ and $3$, respectively. For $k=2$, the map is biholomorphic on $\mathbb{C}\setminus\{0\}$. For $k=3$, the map fails to be bijective outside $\mathbb{D}$ due to the presence of zeros at the cubic roots of $\bar{a}/b$.}
\label{Fig:Branchcut}
\end{figure}

Let us illustrate the points above with flower deformations, for which the function $f(\varphi)$ in Eq.~\eqref{s16b} is given by
\beq
\label{e65}
\ee^{\ii k f(\varphi)}
= \frac{a\, \ee^{\ii k\varphi} + b}
    {\bar{b}\, \ee^{\ii k \varphi} + \bar{a}},
\eeq
where $k$ is some positive integer and $a,b$ are complex coefficients such that $|a|^2-|b|^2=1$. [At fixed $k$, the set of transformations \eqref{e65} spans a group locally isomorphic to $\text{SU}(1,1)$.] The corresponding inverse map \eqref{s15t} is
\beq
\label{eq:gQHinv_kFlower_ab}
\symplectic_{\partial}^{-1}(\ee^{\ii \theta})
= \ee^{\ii\theta}
  \bigl[-\bar{a} + b \ee^{-\ii k\theta}\bigr]^{\frac{k+2}{2k}}
  \bigl[-a + \bar{b} \ee^{\ii k\theta}\bigr]^{\frac{k-2}{2k}},
\eeq
and its unique local conformal extension \eqref{t15t} is
\beq
\label{e67}
H^{-1}(\zeta)
= \zeta
  \bigl[-\bar{a} + b \zeta^{-k}\bigr]^{\frac{k+2}{2k}}
  \bigl[-a + \bar{b} \zeta^{k}\bigr]^{\frac{k-2}{2k}}.
\eeq
Note the simplification of the right-hand side when $k = 2$, corresponding to an elliptic droplet whose biholomorphism was written in Eq.~\eqref{e60}, with $a=-\cosh\lambda$ and $b=\sinh\lambda$. In that case, the answer to Question~\ref{ss16b} is positive: $H^{-1} = \conformal^{-1}$ maps the outside of a disk on the outside of $\droplet$ in a smooth, bijective and conformal manner; see Fig.~\ref{Fig:Branchcut}.

The situation is completely different for flowers with $k\neq2$ petals, since the function \eqref{e67} then has poles (if $k=1$) or zeros (if $k\geq3$) outside $\mathbb{D}$. Any such point means that $H^{-1}$ fails to be a biholomorphism outside $\mathbb{D}$; see Fig.~\ref{Fig:Branchcut} for an illustration when $k=3$. Equivalently, there exists no CG whose shape is that of an edge-deformed flower with $k\neq2$ petals and whose conformal angle is given by Eq.~\eqref{e65}.

\subsection*{Quantum Hall correlations from a local conformal extension}

Despite the general failure of the local conformal map \eqref{t15t} to extend to a global biholomorphism outside $\droplet$, it does have one successful application: it can be used to predict the edge correlations of some anisotropic, $\nu=1$ QH droplets, from the correlator of a chiral, free-fermion Euclidean CFT. Let us show this here, then discuss why it works.

Our starting point is the correlator of chiral free fermions on the complex plane in Euclidean CFT, $K_0(z_1,z_2) = {1}/{(z_1-z_2)}$. Now apply to this a conformal transformation $H$ to find the deformed correlator
\beq
K(z_1,z_2)
=
\frac{\sqrt{H'(z_1)H'(z_2)}}{H(z_1)-H(z_2)},
\label{s16q}
\eeq
where the square root in the numerator arises because the fermionic field is Virasoro primary with conformal weight $1/2$. If $H$ is the conformal extension of the boundary map \eqref{t16b} associated with an edge deformation, then the value of $H(z)$ right on the edge is $H\big(\ee^{\ii f^{-1}(\theta)}/\sqrt{(f^{-1})'(\theta)}\big)=\ee^{\ii\theta}$ by virtue of the map \eqref{s15t}. Furthermore, the derivatives of $H$ in the square root factors of Eq.~\eqref{s16q} have the complex norm
\beq
\left|H'\left(\frac{\ee^{\ii f^{-1}(\theta)}}{\sqrt{(f^{-1})'(\theta)}}\right)\right|
=
\frac{1}{\left|(H^{-1})'(\ee^{\ii\theta})\right|}
\frac{1}{\left|(f^{-1})'(\theta)^{1/2}+\frac{\ii}{2}\frac{(f^{-1})''(\theta)}{(f^{-1})'(\theta)^{3/2}}\right|}
=
\frac{1}{\sqrt{(f^{-1})'(\theta)+\frac{1}{4}\frac{(f^{-1})''(\theta)^2}{(f^{-1})'(\theta)^3}}},
\eeq
where we used Eq.~\eqref{t15t} in the second step. Note that this can also be expressed in a way that makes explicit contact with the notation in \cite{OLMSE:2024}, namely
\beq
\left|H'\left(\frac{\ee^{\ii f^{-1}(f(\varphi))}}{\sqrt{(f^{-1})'(f(\varphi))}}\right)\right|
=
\left|H'\left(\sqrt{f'(\varphi)}\ee^{\ii\varphi}\right)\right|
=
\sqrt{\frac{f'(\varphi)}{1+\left[ \frac{f''(\varphi)}{2f'(\varphi)} \right]^2}}
\equiv
\frac{1}{\sigma(\varphi)}
\eeq
in terms of the polar angle $\varphi$ in the edge deformation \eqref{s16b}. It follows that the norm of the correlator \eqref{s16q}, evaluated right on the edge and expressed in $\varphi$, is
\beq
\big|K(z_1,z_2)\big|_{\textrm{edge}}
=
\frac{1}{\sqrt{\sigma(\varphi_1)\sigma(\varphi_2)}}\frac{1}{2\bigl|\sin\bigl([f(\varphi_1)-f(\varphi_2)]/2\bigr)\bigr|}.
\label{s15q}
\eeq
This is precisely the QH edge correlator \eqref{cedge}, evaluated right on the edge ($d_{\bx}=d_{\by}=0$) of a droplet in an edge-deformed potential \eqref{s1e} with $\theta(\varphi) \equiv f(\varphi)$. Indeed, the norm \eqref{v_theta} of the edge velocity in that case is $v(\theta)=\sqrt{2N} \ell \omega_{\textrm{F}} \sigma(f^{-1}(\theta))$, where $\omega_{\textrm{F}}\equiv\dot{\theta}(\action_N)$ is given by Eq.~\eqref{s15s}.

We have thus recovered a QH correlator \eqref{cedge}, including its delicate dependence on velocities, from an argument involving conformal maps. How does this not contradict our main statement---that QH physics is symplectic while only CGs are conformal?

It is indeed true that the edge correlator of a CG takes the form \eqref{s16q} provided $H(z)$ is replaced by the biholomorphism $\conformal(z)$ in \eqref{ss15n} \cite{HedenmalmWennman:2021, AmeurCronvall:2023}. In this sense, CG correlations and correlations in an edge-deformed QH droplet are close, but there is again a distinction between the maps involved: the map $\conformal$ in \eqref{ss15n} is a biholomorphism outside $\droplet$, while the map $H$ in Eq.~\eqref{t15t} typically is not. As explained above, $H$ and $\conformal$ are generally unrelated, so the QH correlator \eqref{s15q} ultimately has nothing to do with the CG correlator obtained by taking Eq.~\eqref{s16q} and replacing $H$ by $\conformal$. This is the difference between the angles \eqref{s1da}--\eqref{s1db}, all over again. That the conformal ``recipe'' \eqref{s16q} works for edge-deformed droplets is ultimately a coincidence, due to the relation between edge deformations and Virasoro transformations of the edge CFT; see \cite{OLMSE:2024} and references therein.


\bigskip

\noindent {\bf References}

{\small%
\begin{enumerate}[leftmargin=2.0em, itemsep=0.0em, label={[\arabic*]}, ref={\arabic*}, itemindent=0.0em]

\setcounter{enumi}{4}
\item
\label{Laughlin:1983anomalous_SM}
R.~B.~Laughlin,
Anomalous quantum Hall effect: An incompressible quantum fluid with fractionally charged excitations,
\href{https://doi.org/10.1103/PhysRevLett.50.1395}{Phys.\ Rev.\ Lett.\ {\bf 50}, 1395 (1983)}.

\setcounter{enumi}{7}
\item
\label{Jancovici:1995_SM}
B.~Jancovici,
Classical Coulomb systems: Screening and correlations revisited,
\href{https://doi.org/10.1007/BF02178367}{J.\ Stat.\ Phys.\ {\bf 80}, 445 (1995)},
\href{https://arxiv.org/abs/cond-mat/9503109}{arXiv:cond-mat/9503109}.

\setcounter{enumi}{11}
\item
\label{LebleSerfaty:2018_SM}
T.~Lebl\'{e} and S.~Serfaty,
Fluctuations of two dimensional Coulomb gases,
\href{https://doi.org/10.1007/s00039-018-0443-1}{Geom.\ Funct.\ Anal.\ {\bf 28}, 443 (2018)},
\href{https://arxiv.org/abs/1609.08088}{arXiv:1609.08088}.

\setcounter{enumi}{12}
\item
\label{Jancovici:1982ii_SM}
B.~Jancovici,
Classical Coulomb systems near a plane wall. II,
\href{https://doi.org/10.1007/BF01020786}{J.\ Stat.\ Phys.\ {\bf 29}, 263 (1982)}.

\setcounter{enumi}{13}
\item
\label{AlastueyJancovici:1984_SM}
A.~Alastuey and B.~Jancovici,
On potential and field fluctuations in two-dimensional classical charged systems,
\href{https://doi.org/10.1007/BF01018558}{J.\ Stat.\ Phys.\ {\bf 34}, 557 (1984)}.

\setcounter{enumi}{14}
\item
\label{ChoquardEtAl:1987_SM}
Ph.~Choquard, B.~Piller, and R.~Rentsch,
On the dielectric susceptibility of classical Coulomb systems. II,
\href{https://doi.org/10.1007/BF01013377}{J.\ Stat.\ Phys.\ {\bf 46}, 599 (1987)}.

\setcounter{enumi}{27}
\item
\label{OLMSE:2024_SM}
B.~Oblak, B.~Lapierre, P.~Moosavi, J.-M.~St\'{e}phan, and B.~Estienne, Anisotropic quantum Hall droplets,
\href{https://doi.org/10.1103/PhysRevX.14.011030}{Phys.\ Rev.\ X {\bf 14}, 011030 (2024)},
\href{https://arxiv.org/abs/2301.01726}{arXiv:2301.01726}.

\setcounter{enumi}{36}
\item
\label{Fock:1928_SM}
V.~Fock,
Bemerkung zur Quantelung des harmonischen
Oszillators im Magnetfeld,
\href{https://doi.org/10.1007/BF01390750}{Z.\ Physik {\bf 47}, 446 (1928)}.

\setcounter{enumi}{37}
\item
\label{Darwin:1931_SM}
C.~G.~Darwin,
The diamagnetism of the free electron,
\href{https://doi.org/10.1017/S0305004100009373}{Math.\ Proc.\ Camb.\ Philos.\ Soc.\ {\bf 27}, 86 (1931)}.

\setcounter{enumi}{41}
\item
\label{Charles:2003a_SM}
L.~Charles,
Berezin-Toeplitz operators, a semi-classical approach,
\href{https://doi.org/10.1007/s00220-003-0882-9}{Commun.\ Math.\ Phys.\ {\bf 239}, 1 (2003)}.

\setcounter{enumi}{42}
\item
\label{Charles:2003b_SM}
L.~Charles,
Quasimodes and Bohr-Sommerfeld conditions for the Toeplitz operators, 
\href{https://doi.org/10.1081/PDE-120024521}{Commun.\ Partial Differ.\ Equ.\ {\bf 28}, 1527 (2003)}.

\setcounter{enumi}{44}
\item
\label{Dyson:1962_SM}
F.~J.~Dyson,
Statistical theory of the energy levels of complex systems. II,
\href{https://doi.org/10.1063/1.1703774}{J.\ Math.\ Phys.\ {\bf 3}, 157 (1962)}.

\setcounter{enumi}{47}
\item
\label{HedenmalmWennman:2021_SM}
H.~Hedenmalm and A.~Wennman,
Planar orthogonal polynomials and boundary universality in the random
normal matrix model,
\href{https://doi.org/10.4310/ACTA.2021.v227.n2.a3}{Acta Math.\ {\bf 227}, 309 (2021)},
\href{https://arxiv.org/abs/1710.06493}{arXiv:1710.06493}.

\setcounter{enumi}{48}
\item
\label{AmeurCronvall:2023_SM}
Y.~Ameur and J.~Cronvall,
Szeg\H{o} type asymptotics for the reproducing kernel in spaces of full-plane weighted polynomials,
\href{https://doi.org/10.1007/s00220-022-04539-y}{Commun.\ Math.\ Phys. {\bf 398}, 1291 (2023)},
\href{https://arxiv.org/abs/2107.11148}{arXiv:2107.11148}.

\setcounter{enumi}{51}
\item
\label{SaffTotik:1997_SM}
E.~B.~Saff and V.~Totik,
\href{https://doi.org/10.1007/978-3-662-03329-6}{Logarithmic Potentials with External Fields}
(Springer, Berlin, Heidelberg, 1997).

\setcounter{enumi}{52}
\item
\label{DriscollTrefethen:2002_SM}
T.~A.~Driscoll and L.~N.~Trefethen,
\href{https://doi.org/10.1017/CBO9780511546808}{Schwarz-Christoffel Mapping},
Cambridge Monographs on Applied and Computational Mathematics
(Cambridge University Press, Cambridge, 2002).

\setcounter{enumi}{53}
\item
\label{ZabrodinWiegmann:2006_SM}
A.~Zabrodin and P.~Wiegmann,
Large-$N$ expansion for the 2D Dyson gas,
\href{https://doi.org/10.1088/0305-4470/39/28/S10}{J.\ Phys.\ A: Math.\ Gen.\ {\bf 39}, 8933 (2006)},
\href{https://arxiv.org/abs/hep-th/0601009}{arXiv:hep-th/0601009}.

\setcounter{enumi}{64}
\item
\label{CardosoEtAl:2021_SM}
G.~Cardoso, J.-M.~St\'{e}phan, and A.~G.~Abanov, 
The boundary density profile of a Coulomb droplet. Freezing
at the edge,
\href{https://doi.org/10.1088/1751-8121/abcab9}{J.\ Phys.\ A: Math.\ Theor.\ {\bf 54}, 015002 (2020)},
\href{https://arxiv.org/abs/2009.02359}{arXiv:2009.02359}.

\setcounter{enumi}{65}
\item
\label{StillingerLovett:1968_SM}
F.~Stillinger and R.~Lovett,
Ion-pair theory of concentrated electrolytes. I. Basic concepts,
\href{https://doi.org/10.1063/1.1669709}{J.\ Chem.\ Phys.\ {\bf 48}, 3858 (1968)}.

\end{enumerate}
}

\end{document}